\documentclass{aastex}
\usepackage{emulateapj5}
\submitted{Received 2003 January 22; accepted 2003 June 4}
\journalinfo{THE ASTROPHYSICAL JOURNAL, 595, 2003 September 20}
\shortauthors{SAIJO, BAUMGARTE, {\&} SHAPIRO}
\shorttitle{ONE-ARMED SPIRAL INSTABILITY IN DIFFERENTIALLY
ROTATING STARS}
\begin{document}
%
\title{One-Armed Spiral Instability in
Differentially Rotating Stars}
\author{Motoyuki Saijo}
\email{saijo@tap.scphys.kyoto-u.ac.jp}
\affil
{Department of Physics, Kyoto University,
Kyoto 606-8502, Japan}
%
\author{Thomas W. Baumgarte\altaffilmark{1}}
\email{tbaumgar@bowdoin.edu}
\affil
{Department of Physics and Astronomy, Bowdoin College,\\
8800 College Station
Brunswick, ME 04011}
\altaffiltext{1}{
Department of Physics, 
University of Illinois at Urbana-Champaign,
Urbana, IL 61801}
%
\author{Stuart L. Shapiro\altaffilmark{2,3}}
\email{shapiro@astro.physics.uiuc.edu}
\affil{
Department of Physics, 
University of Illinois at Urbana-Champaign,
Urbana, IL 61801}
\altaffiltext{2}
{Department of Astronomy, 
University of Illinois at Urbana-Champaign,
Urbana, IL 61801}
\altaffiltext{3}
{NCSA, 
University of Illinois at Urbana-Champaign,
Urbana, IL 61801}
%
\begin{abstract}
We investigate the dynamical instability of the one-armed spiral $m=1$
mode in differentially rotating stars by means of $3+1$ hydrodynamical
simulations in Newtonian gravitation.  We find that both a soft
equation of state and a high degree of differential rotation in the
equilibrium star are necessary to excite a dynamical $m=1$ mode as the
dominant instability at small values of the ratio of rotational
kinetic to gravitational potential energy, $T/|W|$.  
We find that this spiral mode
propagates outward from its point of origin near the maximum density
at the center to the surface over several central orbital periods.  An
unstable $m=1$ mode triggers a secondary $m=2$ bar mode of smaller
amplitude, and the bar mode can excite gravitational waves.  As the
spiral mode propagates to the surface it weakens, simultaneously
damping the emitted gravitational wave signal.  This behavior is in
contrast to waves triggered by a dynamical $m=2$ bar instability,
which persist for many rotation periods and decay only after a
radiation-reaction damping timescale.
\end{abstract}
\keywords{Gravitation --- hydrodynamics --- instabilities
 --- stars: neutron --- stars: rotation}
%
\section{Introduction}

Stars in nature are usually rotating and may be subject to
nonaxisymmetric rotational instabilities.  An exact treatment of these
instabilities exists only for incompressible equilibrium fluids in
Newtonian gravity, \citep[e.g.][]{Chandra69,Tassoul}.  For these
configurations, global rotational instabilities may arise from
non-radial toroidal modes $e^{im\varphi}$ (where $m=\pm 1,\pm 2,
\dots$ and $\varphi$ is the azimuthal angle).

For sufficiently rapid rotation, the $m=2$ bar mode becomes either
{\em secularly} or {\em dynamically} unstable.  The onset of
instability can typically be identified with a critical value of the
non-dimensional parameter $\beta \equiv T/|W|$, where $T$ is the
rotational kinetic energy and $W$ the gravitational potential energy.
Uniformly rotating, incompressible stars in Newtonian theory are
secularly unstable to bar-mode formation when $\beta \geq \beta_{\rm
sec} \simeq 0.14$.  This instability can grow only in the presence of
some dissipative mechanism, like viscosity or gravitational radiation,
and the associated growth timescale is the dissipative timescale,
which is usually much longer than the dynamical timescale of the
system.  By contrast, a dynamical instability to bar-mode formation
sets in when $\beta \geq \beta_{\rm dyn} \simeq 0.27$.  This
instability is independent of any dissipative mechanisms, and the
growth time is the hydrodynamic timescale.

Determining the onset of the dynamical bar-mode instability, as well
as the subsequent evolution of an unstable star, requires a fully
nonlinear hydrodynamic simulation.  Simulations performed in Newtonian
gravity, \citep[e.g.][]{TDM,DGTB,WT,HCS,SHC,HC,PDD,TIPD,NCT} have
shown that $\beta_{\rm dyn}$ depends only very weakly on the stiffness
of the equation of state.  Once a bar has developed, the formation of
a two-arm spiral plays an important role in redistributing the angular
momentum and forming a core-halo structure.  Both $\beta_{\rm dyn}$
and $\beta_{\rm sec}$ are smaller for stars with high degree of
differential rotation \citep{TH,PDD,SKE02,SKE03}.  Simulations in
relativistic gravitation \citep{SBS00,SSBS} have shown that
$\beta_{\rm dyn}$ decreases with the compaction of the star,
indicating that relativistic gravitation enhances the bar mode
instability.  In order to efficiently use computational resources,
most of these simulations have been performed under certain symmetry
assumptions ({\it e.g.}  $\pi$-symmetry), which do not affect the
growth of the $m=2$ bar mode, but which suppress any $m=1$ modes.

Recently, \citet{CNLB} reported that such $m=1$ ``one-armed spiral''
modes are dynamically unstable at surprisingly small values of
$T/|W|$.  \citet{CNLB} found this instability in evolutions of highly
differentially rotating equilibrium polytropes with polytropic index
$n=3.33$.  Typically, these equilibria have a ``toroidal'' structure, 
so that the maximum density is not located at the geometric center but 
rather on a toroid rotating about the center.

It is possible that the $m=1$ instability in equilibrium stars is
related to that arising in protostellar disk systems.  This
instability originally was found in nearly Keplerian, thin, gaseous
disks around central point masses, both numerically \citep{ARS,HPS}
and analytically \citep{STAR}.  The central point mass moves away from
the center of mass of the whole system due to a perturbation and this
displacement triggers the instability.  This particular mode of
instability only occurs when the mass ratio $M_{\rm disk}/M_{\rm
total}$ exceeds $0.2$.  An $m=1$ instability has also been found in
thick, self-gravitating, protostellar tori \citep{WTH} and
protostellar disks \citep{LB}, as well as in finite fluid cores
surrounded by disk halos \citep{PDD}.  In the latter case, the
instability arises from the internal interaction between different
regions of a single, continuous body, and the disk does not need to
satisfy the above mass criterion to trigger the $m=1$ instability
\citep[see also][for an example of an unstable central accreting
object surrounded by a rotationally supported gas disk]{Bonnell}.

The purpose of this paper is to study further the conditions under
which a dynamical $m=1$ instability is excited.  We vary both the
polytropic index, i.e.~the stiffness of the equation of state, and the
degree of differential rotation to isolate their effects on the
instability.  Since the onset of rotational
instabilities is often characterized by $\beta$ we keep this value
approximately fixed in our sequences.  We find that a soft equation of
state and a high degree of differential rotation are both necessary to
dynamically excite the $m=1$ mode at the small value of $\beta = 0.14$
chosen in this paper.  We find that a toroidal structure is not
sufficient to trigger the $m=1$ instability, but our findings suggest
that a toroidal structure may be necessary.

While our goal is to gain a deeper understanding of the nature of the
$m=1$ instability as opposed to simulating realistic astrophysical
scenarios, we point out that there exist evolutionary sequences that
may well lead to rapidly and highly differentially rotating
configurations.  For example, cooling by thermal emission from a
rotating star will cause the star to contract and spin up.  If
internal viscosity and magnetic fields are sufficiently weak, this
process will lead to differential rotation even if the initial
configuration is rotating uniformly.  This scenario may arise in
supermassive stars, where the equation of state is dominated by
radiation pressure and may be modeled by a (soft) $n=3$ polytrope.  In
the absence of viscosity and magnetic braking, the star will contract
quasi-statically as it cools to a toroidal configuration, which may be
subject to $m=1$ or $m=2$ instabilities \citep{NS}.  
Stellar collisions and mergers may also lead to differentially rotating stars.
For the coalescence of binary neutron stars \citep{SU00,SU02}, the
presence of differential rotation may temporarily stabilize the
``hypermassive'' remnant and may therefore have important dynamical
effects \citep{BSS00,LBS03}.  However, as we find in this paper, the
$m=1$ mode is unstable only for very soft equations of state, so that
it is not obvious that they will arise in the remnant of a binary neutron star
merger. However, they may arise in a rapidly spinning proto-neutron star
core when surrounded by a fall-back disk, possibly forming a low mass 
condensation which can explode and induce a large neutron star 
recoil speed \citep{CW}. Finally, the $m=1$ instability might arise in
massive disks around black holes, especially if the disks are 
radiation-dominated, and hence governed by a soft equation of state.

This paper is organized as follows.  In \S~\ref{sec:Nhydro} we present
the basic equations, our initial data and diagnostics.  We discuss our
numerical results in \S~\ref{sec:Dynamics}, and briefly summarize our
findings in \S~\ref{sec:Discussion}.  Throughout this paper we use
gravitational units \footnote{Since we adopt Newtonian gravity in this
paper, the speed of light only enters in the gravitational waveforms
(\S~\ref{subsec:GW} and \S~\ref{sec:Dynamics}).} with $G = c = 1$ and
adopt Cartesian coordinates $(x,y,z)$.

\section{Basic Equations}
\label{sec:Nhydro}
\subsection{Newtonian Hydrodynamics}
We construct a $3+1$ dimensional Newtonian hydrodynamics code assuming
an adiabatic $\Gamma$-law equation of state
\begin{equation}
P = ( \Gamma - 1 ) \rho \varepsilon,
\label{eqn:GammaLaw}
\end{equation}
where $P$ is the pressure, $\Gamma$ the adiabatic index, $\rho$ the
mass density and $\varepsilon$ the specific internal energy density.
For perfect fluids the Newtonian equations of hydrodynamics then
consist of the continuity equation
\begin{equation}
\frac{\partial \rho}{\partial t}
+\frac{\partial (\rho v^{i})}{\partial x^{i}} = 0,
\label{eqn:continuity}
\end{equation}
the energy equation
\begin{equation}
\frac{\partial e}{\partial t}+
\frac{\partial (e v^{j})}{\partial x^{j}} = 
- \frac{1}{\Gamma} e^{-(\Gamma-1)} P_{\rm vis} 
\frac{\partial v^{i}}{\partial x^{i}}
,
\end{equation}
and the Euler equation
\begin{equation}
\frac{\partial(\rho v_{i})}{\partial t}
+ \frac{\partial (\rho v_{i} v^{j})}{\partial x^{j}} 
=
- \frac{\partial (P + P_{\rm vis})}{\partial x^{i}}
- \rho \frac{\partial \Phi}{\partial x^{i}}.  
\end{equation}
Here $v^i$ is the fluid velocity, $\Phi$ is the gravitational potential, 
and $e$ is defined according to
\begin{equation}
e = (\rho \varepsilon)^{1/\Gamma}.
\end{equation}
We compute the artificial viscosity pressure $P_{\rm vis}$ from \citep{RM}
\begin{equation}
P_{\rm vis} =
\cases{ 
C_{\rm vis} 
\rho (\delta v)^{2},
& for $\delta v \leq 0$;
\cr
0, & for $\delta v \geq 0$,\cr
}
\end{equation}
where $\delta v \equiv 2 \delta x \partial_{i} v^{i}$, $\delta x (=
\Delta x = \Delta y = \Delta z)$ is the local grid spacing and where
we choose the dimensionless parameter $C_{\rm vis} = 2$.  When
evolving the above equations we limit the stepsize $\Delta t$ by an
appropriately chosen Courant condition.

We have tested the ability of our code to resolve shocks by performing
a wall-shock problem, in which two phases of a fluid collide at
supersonic speeds.  In Fig.~\ref{fig:wshock} we compare numerical
results with the analytic solution for initial velocities that are
similar to those found in our simulations below.  With $C_{\rm vis} =
2$ we find good agreement for Mach numbers up to $M_{\rm mach}
\lesssim 6$.  The drop in density at $x=0$ is usually interpreted as
``wall heating'' \citep[e.g.][]{HSW}.

\begin{center}
\begin{minipage}{7.0cm}
\epsfxsize 7.0cm
\epsffile{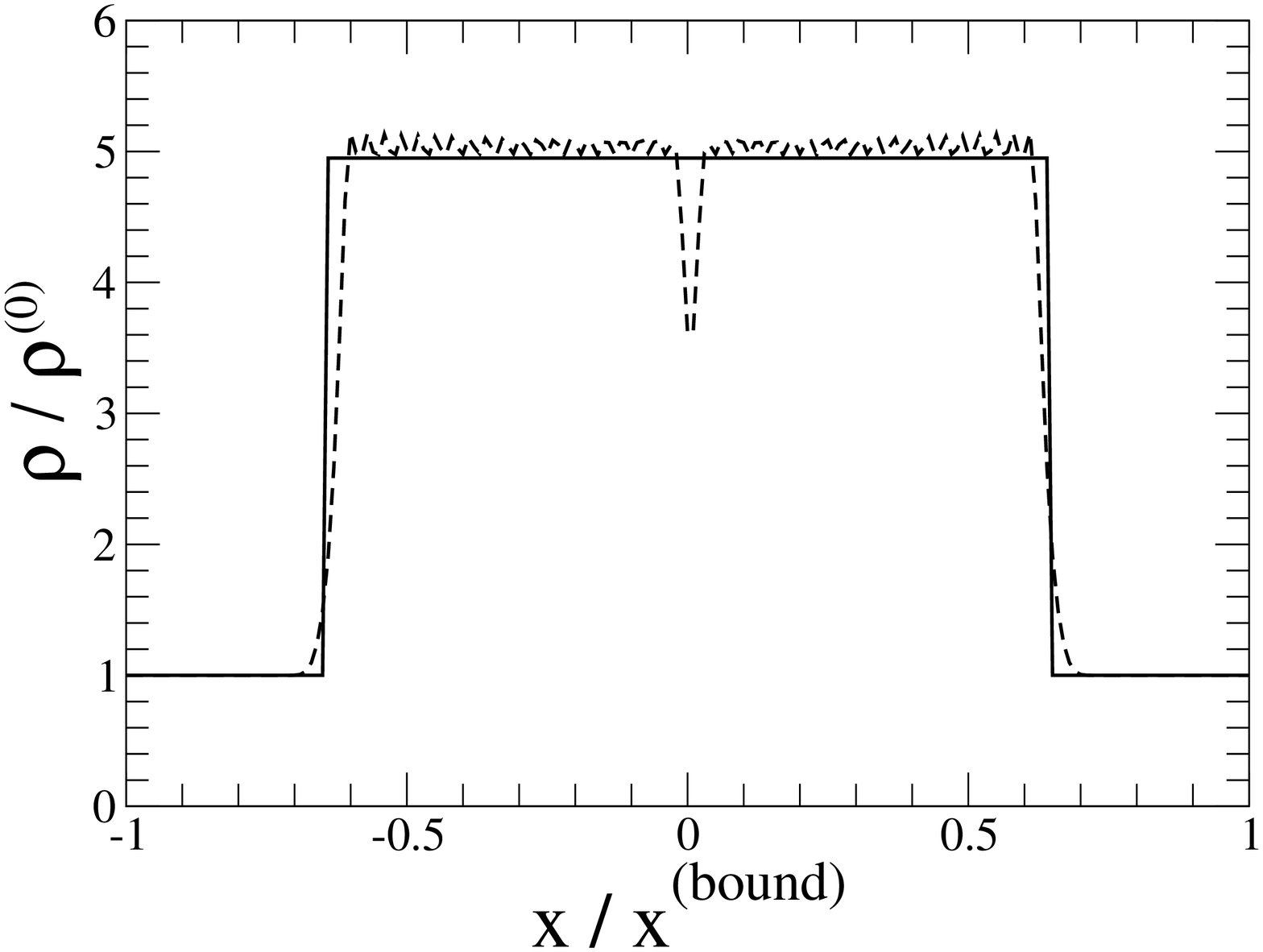}
\end{minipage}
\end{center}
\figurenum{1}
\figcaption[f1.eps]{
Comparison between numerical and analytical results in a
one-dimensional wall shock problem at $t=2.5 x^{\rm (bound)} / v_{0}$
(where the fluid flow is aligned with the $x$-axis).  Solid and dashed
lines represent analytic and numerical results, respectively.  For
this simulation we chose $\Gamma = 1.30$, $\rho^{(0)} = 2.80 \times
10^{-3}$, $\kappa = 5.85 \times 10^{-2}$, $x^{\rm (bound)} = 0.5$ with
a grid space $\delta x = 5 \times 10^{-3}$ and $v_{0} = 2.78 v_{s}$,
where $v_{s}$ is the initial speed of sound.
\label{fig:wshock}
}
\vskip 12pt

The gravitational potential is determined by the Poisson equation
\begin{equation}
\triangle \Phi = 4 \pi \rho,
\end{equation}
which we solve subject to the outer boundary condition
\begin{equation}
\Phi = - \frac{M}{r} - \frac{d_{i} x^{i}}{r^{2}} + O(r^{-3}).
\end{equation}
Here $M$ is the total mass
\begin{equation}
M = \int_{V} \rho dx^{3}
\label{eqn:M}
\end{equation}
and $d_i$ is the dipole moment
\begin{equation}
d_{i} = \int_{V} \rho x_{i} dx^{3}.
\end{equation}

\subsection{Initial Data}
As initial data, we construct differentially rotating equilibrium models
with an algorithm based on \citet{Hachisu}.  Individual models are
parameterized by the ratio of the polar to equatorial radius $R_{\rm
p}/R_{\rm eq}$, and a parameter of dimension length $d$ that
determines the degree of differential rotation through
\begin{equation}
\Omega = \frac{j_{0}}{d^{2} + \varpi^{2}}.
\label{eqn:omega}
\end{equation}
Here $\Omega$ is the angular velocity, $j_{0}$ is a constant parameter
with units of specific angular momentum, and $\varpi$ is the
cylindrical radius.  The parameter $d$ determines the length scale
over which $\Omega$ changes; uniform rotation is achieved in the limit
$d \rightarrow \infty$.  For the construction of initial data we also
assume a polytropic equation of state
\begin{equation}
P = \kappa \rho^{1+1/n},
\end{equation}
where $n=1/(\Gamma-1)$ is the polytropic index and $\kappa$ a
constant.  In absence of shocks, the polytropic form of the equation
of state is conserved by the $\Gamma$-law equation of state (eq.
(\ref{eqn:GammaLaw})).

To enhance any $m=1$ or $m=2$ instability, we disturb the initial
equilibrium density $\rho_{\rm eq}$ by a non-axisymmetric perturbation
according to\footnote{The numerical finite difference error is in
principle sufficient to trigger instabilities, but starting from
such a small amplitude it would take the instability prohibitively
long to reach saturation.}
\begin{equation}
\rho = \rho_{\rm eq}
\left( 1 + 
  \delta^{(1)} \frac{x+y}{R_{\rm eq}} +
  \delta^{(2)} \frac{x^{2}-y^{2}}{R_{\rm eq}^{2}}
\right),
\label{eqn:DPerturb}
\end{equation}
with $\delta^{(1)} = \delta^{(2)} = 10^{-3}$ in all our simulations.

\subsection{Gravitational Waveforms}
\label{subsec:GW}
We compute approximate gravitational waveforms by evaluating the
quadrupole formula.  In the radiation zone, gravitational waves can be
described by a transverse-traceless, perturbed metric $h_{ij}^{TT}$
with respect to a flat spacetime. In the quadrupole formula,
$h_{ij}^{TT}$ is found from \citep{MTW}
\begin{equation}
h_{ij}^{TT}= \frac{2}{r} \frac{d^{2}}{d t^{2}} I_{ij}^{TT},
\label{eqn:wave1}
\end{equation}
where $r$ is the distance to the source, $I_{ij}$ the quadrupole
moment of the mass distribution \citep[see eq.~{[36.42b]} in][]{MTW},
and where $TT$ denotes the transverse-traceless projection.  Choosing
the direction of the wave propagation to be along the $z$ axis, the two
polarization modes of gravitational waves can be determined from
\begin{equation}
h_{+} \equiv \frac{1}{2} (h_{xx}^{TT} - h_{yy}^{TT})
\mbox{~~~and~~~} 
h_{\times} \equiv h_{xy}^{TT}.
\end{equation}
For observers along the $z$-axis, we thus have
\begin{eqnarray}
\frac{r h_{+}}{M} &=& 
\frac{1}{2 M} \frac{d}{d t} (\dot{I}_{xx} - \dot{I}_{yy}), \label{h+}
\\
\frac{r h_{\times}}{M} &=& 
\frac{1}{M} \frac{d}{d t} \dot{I}_{xy} \label{h-}
.
\end{eqnarray}
The number of time derivatives $I_{ij}$ that have to be taken can be
reduced by using the continuity equation (\ref{eqn:continuity})
\begin{equation}
\dot{I}_{ij} = \int (\rho v^{i} x^{j} + \rho x^{i} v^{j}) d^{3}x,
\end{equation}
in equations (\ref{h+}) and (\ref{h-}) \citep[see][]{Finn}.

\subsection{Numerical Implementation and Diagnostics}
\begin{table*}
\begin{center}
\tablenum{1}
\label{tab:BarTest}
\centerline{\sc Table 1}
\centerline{\sc Initial data for bar formation tests ($n = 1$).}
\vskip 6pt
\begin{tabular}{c c c c c c c c}
\hline
\hline
$d / R_{\rm eq}$ \tablenotemark{a} &
$R_{\rm p} / R_{\rm eq}$ \tablenotemark{b} &
$\Omega_{\rm c} / \Omega_{\rm eq}$ \tablenotemark{c} &
$\rho_{\rm c} / \rho_{\rm max}$ \tablenotemark{d} &
$R_{\rm maxd}/R_{\rm eq}$ \tablenotemark{e} &
$T/|W|$ \tablenotemark{f} &
$m=1$ &
$m=2$
\\
\hline
$0.20$ & $0.250$ & $26.0$ & $0.160$ & $0.383$ &
$0.119$ & Stable & Unstable
\\
\hline
\hline
\end{tabular}
\vskip 12pt
\begin{minipage}{13cm}
${}^{a}$ {$R_{\rm eq}$ = equatorial radius.}
\\
${}^{b}$ {$R_{\rm p}$ = polar radius.}
\\
${}^{c}$ {$\Omega_{\rm c}$ = central angular velocity,
$\Omega_{\rm eq}$ = equatorial angular velocity at the surface.}
\\
${}^{d}$ {$\rho_{\rm c}$ = central density; $\rho_{\rm
max}$ = maximum density.}
\\
${}^{e}$ {$R_{\rm maxd}$ = distance between the origin
and the location of maximum density.}
\\
${}^{f}$ {$T$ = rotational kinetic energy; $W$ =
gravitational potential energy.}
\end{minipage}
\end{center}
\end{table*}

Our code is based on the post-Newtonian hydrodynamics scheme of
\citet{SBS98} and \citet{SSBS}, to which the reader is referred for a
more detailed description, discussion and tests.  We choose the axis
of rotation to align with the $z$ axis, and assume planar symmetry
across the equator.  The equations of hydrodynamics are then solved on
a uniform grid of size $169 \times 169 \times 85$.  We terminate our
simulations either when the central density has increased to a point
at which our resolution becomes inadequate, or after a sufficient
number of central rotation periods (between 20 and 40) in order for us
to detect dynamical instabilities.

We monitor the conservation of mass $M$ (eq.~[\ref{eqn:M}]), angular
momentum $J$
\begin{equation}
J = \int \rho ( x v^{y} - y v^{x} ) d^3x,
\end{equation}
energy $E$
\begin{equation}
E = T + U + W = 
	\frac{1}{2} \int \rho v^{i} v_{i} d^3 x
	+ \int \rho \varepsilon d^3 x
	+ \frac{1}{2} \int \rho \Phi d^3 x,
\label{eqn:NewtonE}
\end{equation}
and the location of the center of mass $x^i_{\rm CM}$
\begin{equation}
x^{i}_{\rm CM} = \int \rho x^{i} d^3 x.
\end{equation}
Here $T$ is the kinetic energy (all rotational at $t=0$), $U$ the
internal energy, and $W$ the gravitational potential energy.  Given
our assumption of equatorial symmetry, we have $x^z_{\rm CM} = 0$
identically, so that we only need to monitor the $x$ and $y$
components of $x^{i}_{\rm CM}$.  Due to our flux-conserving difference
scheme the mass $M$ is also conserved up to round-off error, except if
matter leaves the computational grid (which was never more than 0.01
\% of the total mass).  In all cases reported in \S~\ref{sec:Dynamics}
the energy $E$ and the angular momentum $J$ were conserved up to $\sim
0.1 \%$ of their initial values, and the center of mass moved by less
than about $1 \%$ of one spatial grid cell per central rotation
period.

To monitor the development of $m=1$ and $m=2$ modes we compute a
``dipole diagnostic''\footnote{Our diagnostics differ from those in
previous treatments \citep{CNLB}, where the growth of the mode is
measured at a single, arbitrary Eulerian radius in the equatorial plane inside
the star.}
\begin{equation}
D = \left< e^{i m \varphi} \right>_{m=1} =
\frac{1}{M} \int \rho \frac{x + i y}{\sqrt{x^{2}+y^{2}}} d^3 x
\label{eqn:dipole}
\end{equation}
and a ``quadrupole diagnostic''
\begin{equation}
Q = \left< e^{i m \varphi} \right>_{m=2} =
\frac{1}{M} \int \rho \frac{(x^{2}-y^{2}) + i (2 x y)}{x^{2}+y^{2}} d^3 x,
\label{eqn:quadrupole}
\end{equation}
where a bracket denotes the density weighted average.  In the
following we only plot the real parts of $D$ and $Q$.

\section{Results}
\label{sec:Dynamics}

\subsection{Dynamical bar formation}
\label{subsec:m2test}

Before studying $m=1$ one-armed spiral instabilities, it is useful to
test the capability of our code and our diagnostics to detect any
instabilities.  To do so, we reproduce an $m=2$ bar mode instability
that was recently found by \citet{SKE02} in highly differentially
rotating $n = 1$ polytropes for surprisingly small values of $T/|W|$.
The parameters of our initial data for this test are listed in Table
\ref{tab:BarTest}.  For all our simulations we add a small dipole
($m=1$) and quadrupole ($m=2$) perturbation to the initial equilibrium
star (eq.~[\ref{eqn:DPerturb}]) to enhance the growth of any
instability.
\begin{center}
\begin{minipage}{8.0cm}
\epsfxsize 8.0cm
\epsffile{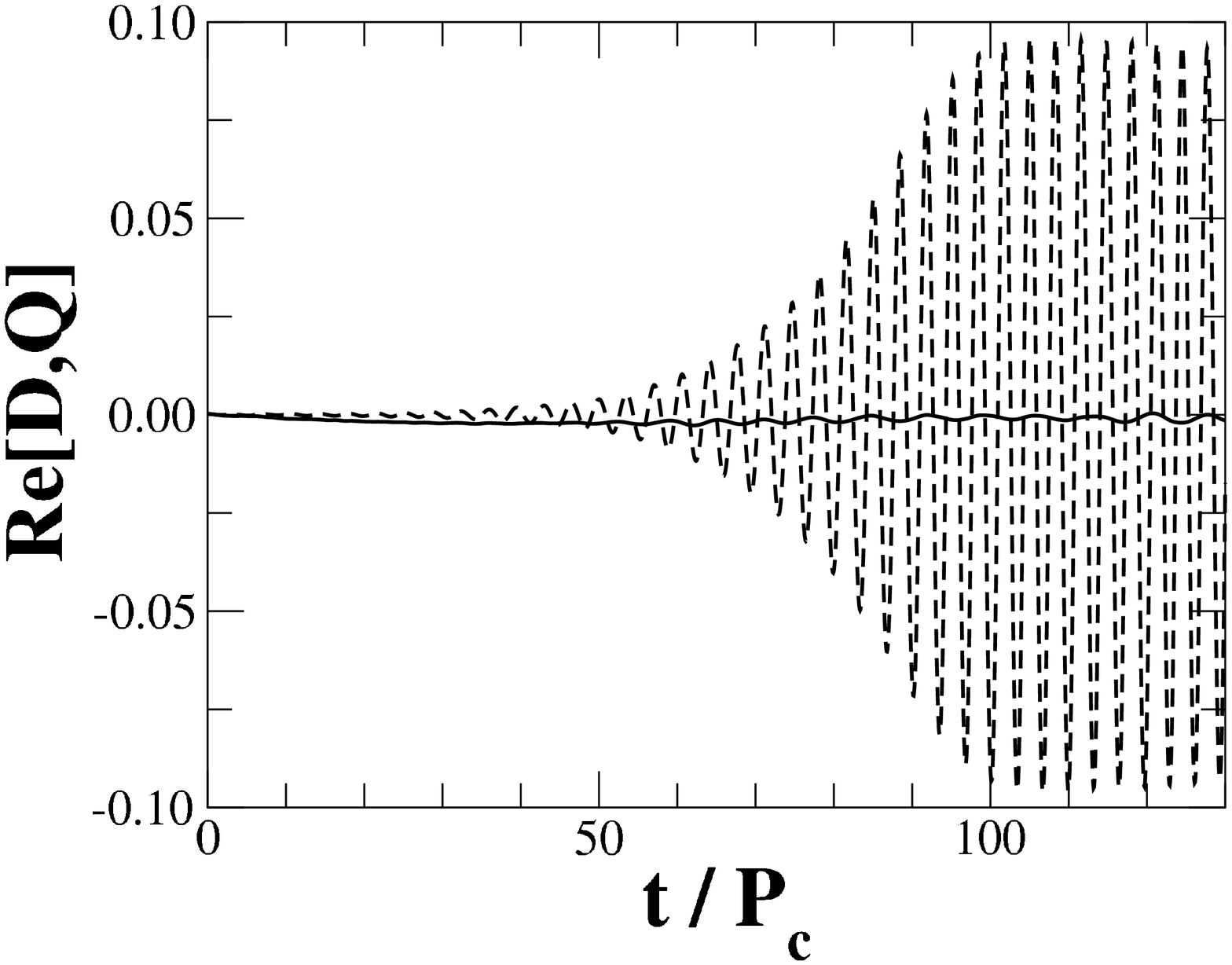}
\end{minipage}
\end{center}
\figurenum{2}
\figcaption[f2.eps]{
Diagnostics $D$ and $Q$ as a function of $t/P_{\rm c}$ for our bar
formation model (see Table \ref{tab:BarTest}).  Solid and dotted lines
denote the values of $D$ and $Q$, respectively.  We terminate our
simulation when $t = 132 P_{\rm c}$.  Hereafter $P_{\rm c}$ represents 
the central rotation period.
\label{fig:bardip}}
\vskip 12pt
\begin{center}
\begin{minipage}{7.5cm}
\epsfxsize 7.5cm
\epsffile{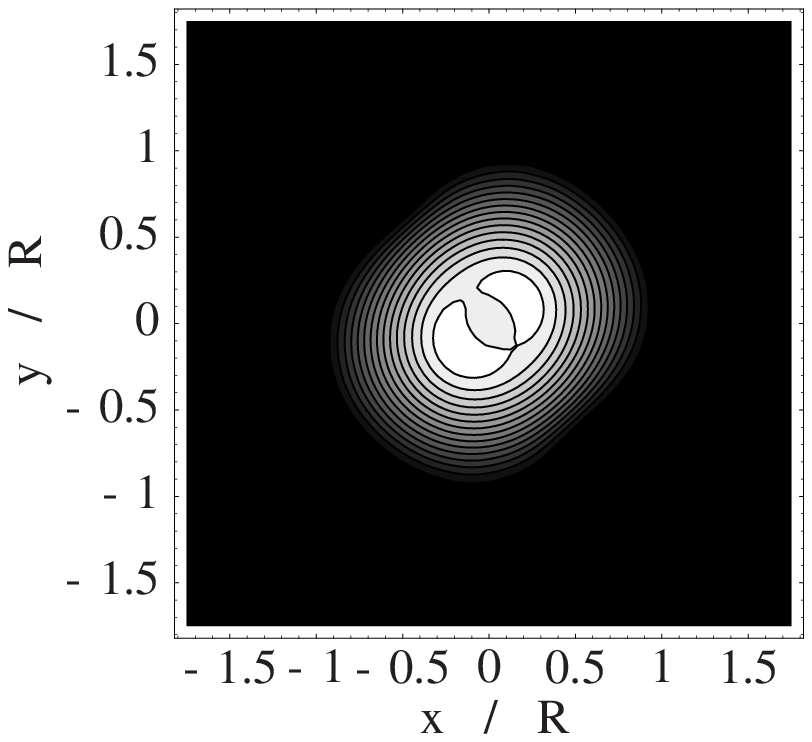}
\end{minipage}
\end{center}
\figurenum{3}
\figcaption[f3.eps]{
Final density contours in the equatorial plane for our bar formation model. 
Snapshots are plotted at ($t/P_{\rm c}$, $\rho_{\rm
max}/\rho_{\rm max}^{(0)}$) = (132, 1.25), where $\rho_{\rm max}$ is 
the maximum density, $\rho_{\rm max}^{(0)}$ is the initial
maximum density, and
$R$ is the initial equatorial radius.  The contour lines 
denote densities $\rho/\rho_{\rm max} = 6.67 \times (16-i) \times
10^{-2} (i=1, \cdots, 15)$. 
\label{fig:barcon}}
\vskip 12pt

\begin{figure*}[b]
\figurenum{5}
\epsscale{1.20}
\plotone{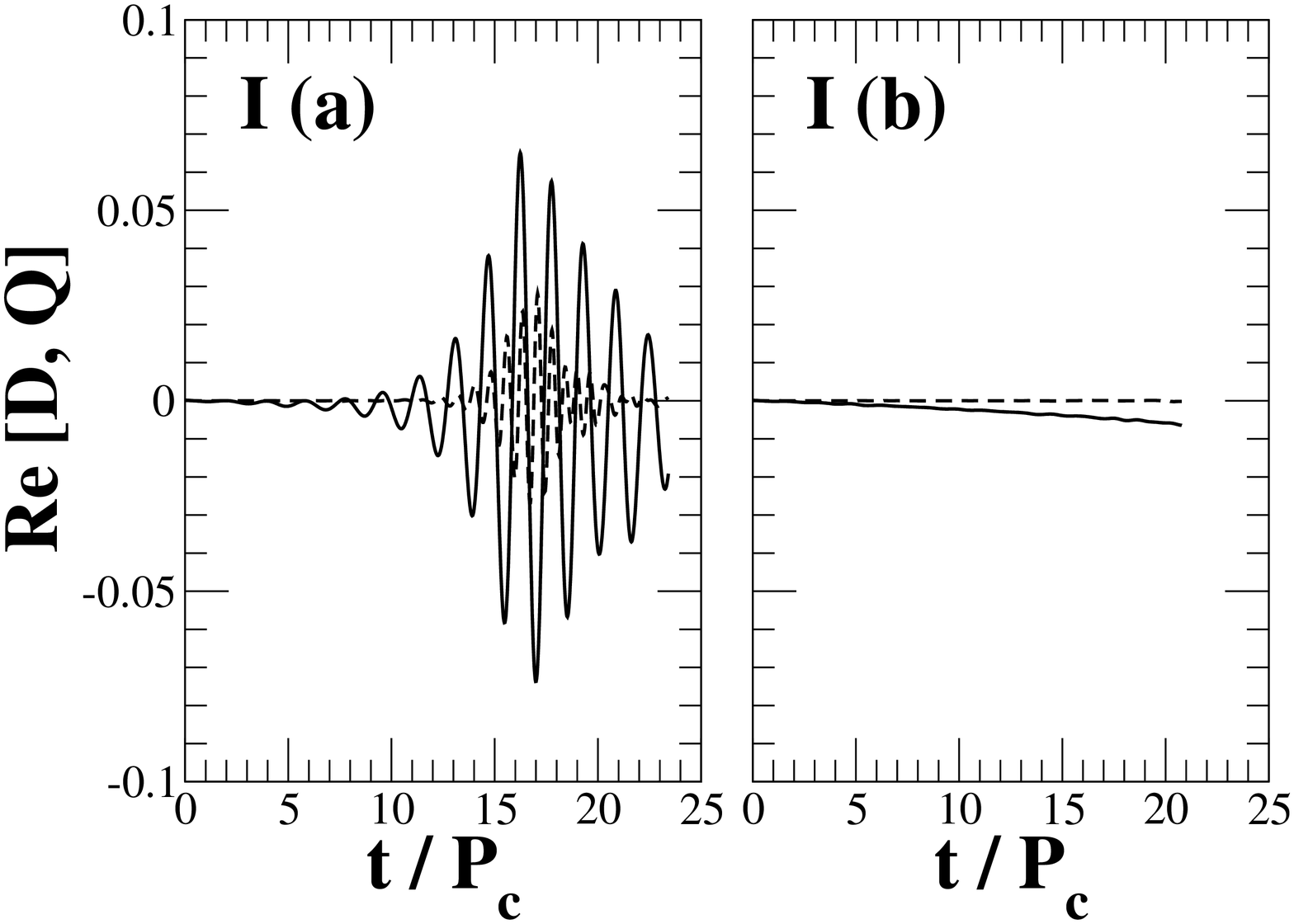}
\figurenum{5}
\figcaption[f5.eps]{
Diagnostics $D$ and $Q$ as a function of $t/P_{\rm c}$ for Model I (a)
and (b) (see Table \ref{tab:m1test}). Solid and dotted lines denote $D$ 
and $Q$.  We terminate our simulation at $t \sim 20 P_{\rm c}$ or 
when the maximum
density of the star exceeds about 10 times its initial value.
\label{fig:cttdip}}
\end{figure*}

\begin{center}
\begin{minipage}{7.0cm}
\epsfxsize 7.0cm
\epsffile{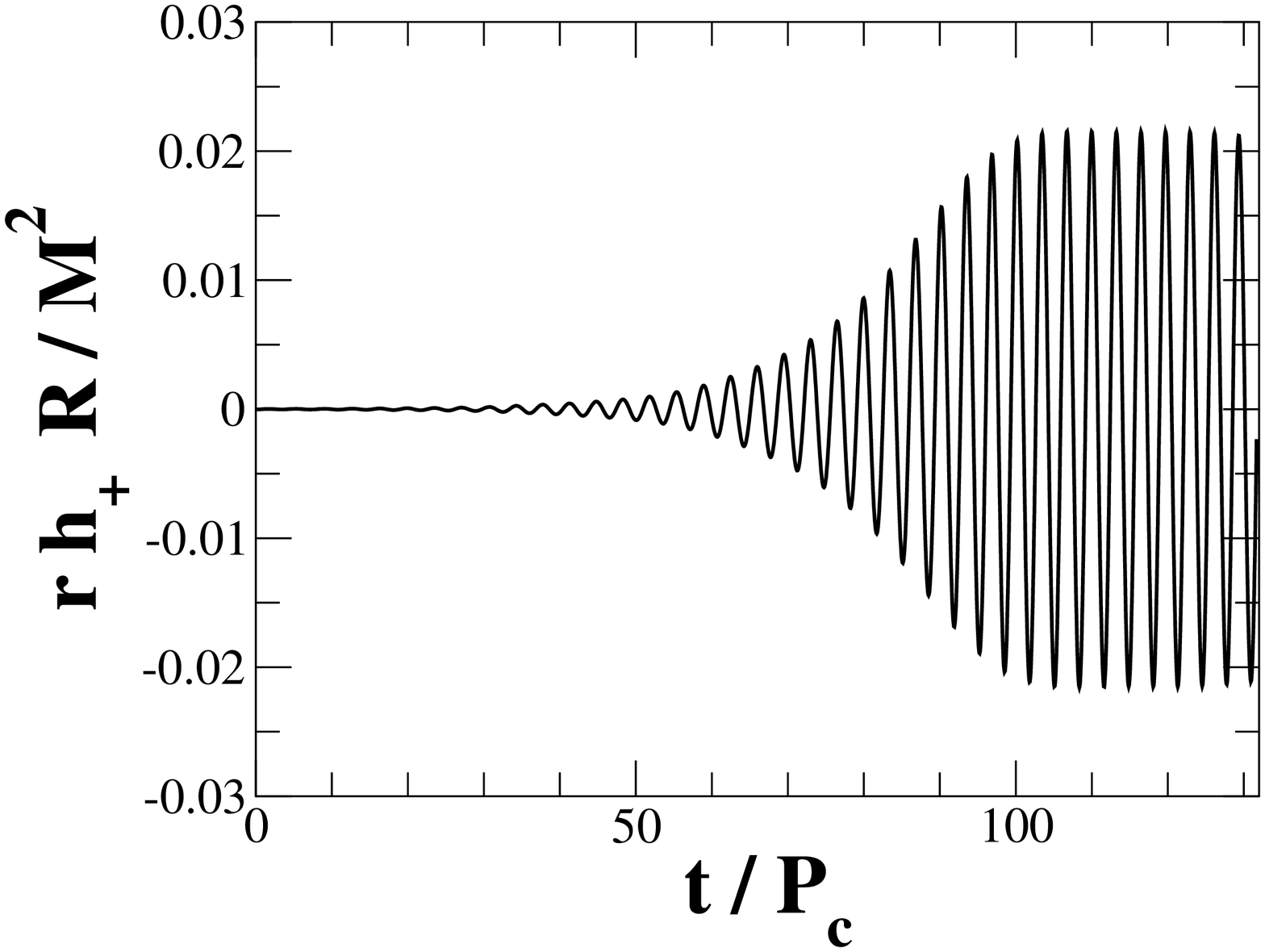}
\end{minipage}
\end{center}
\figurenum{4}
\figcaption[f4.eps]{
Gravitational waveform for bar-unstable star as seen by a distant
observer located on the $z$-axis.
\label{fig:bargw}}
\vskip 12pt

In Fig.~\ref{fig:bardip} we show both diagnostics $D$ and $Q$ as a
function of time.  The dipole diagnostic $D$ remains very small
throughout the evolution (small oscillations are due to the initial
perturbation), while the quadrupole diagnostic $Q$ grows exponentially
until it saturates.  These results indicate that the star is unstable
towards bar formation but stable towards one-armed spiral formation.
The bar persists without decay for over one surface-rotation period
following saturation, corresponding to over 30 central rotation
periods.  After this we terminate our integration.

The bar mode formation is also evident in Fig.~\ref{fig:barcon}, which
shows a snapshot of the density contours just before we terminate the
evolution.  Owing to the small value of $T/|W|$ the bar is too weak to
form double spiral arms.  The gravitational waveform emitted by the
bar formation is shown in Fig.~\ref{fig:bargw}.  We expect that it
will survive without decay until gravitational radiation-reaction
forces destroy the bar ($\sim (R/M)^{5/2} t_{\rm dyn} \gg t_{\rm
dyn}$).

These simulations indicate that our code and diagnostics are capable
of detecting instabilities, and also reconfirm the findings of 
\citet{SKE02} that strongly differentially rotating stars can be
unstable to dynamical bar mode formation even at very small values
of $T/|W|$.

\subsection{Dynamical one-armed spiral formation}
\label{subsec:m1test}

\begin{table*}[t]
\begin{center}
\tablenum{2}
\label{tab:m1test}
\centerline{\sc Table 2}
\centerline{\sc Initial data for the $m=1$ test ($n = 3.33$).}
\vskip 6pt
\begin{tabular}{c c c c c c c c}
\hline
\hline
Model &
$d / R_{\rm eq}$ &
$R_{\rm p} / R_{\rm eq}$ &
$\Omega_{\rm c} / \Omega_{\rm eq}$ &
$\rho_{\rm c} / \rho_{\rm max}$ &
$R_{\rm maxd}/R_{\rm eq}$ &
$T/|W|$ &
$m=1$
\\
\hline
I (a) & $0.20$ & $0.417$ & $26.0$ & $0.531$ & $0.192$ &
$0.144$ & Unstable
\\
I (b) & $0.20$ & $0.542$ & $26.0$ & $1.00$  & $0.00$& 
$0.090$ & Stable
\\
\hline
\hline
\end{tabular}
\end{center}
\vskip 12pt
\end{table*}

\begin{figure*}[t]
\figurenum{6}
\epsscale{1.50}
\plotone{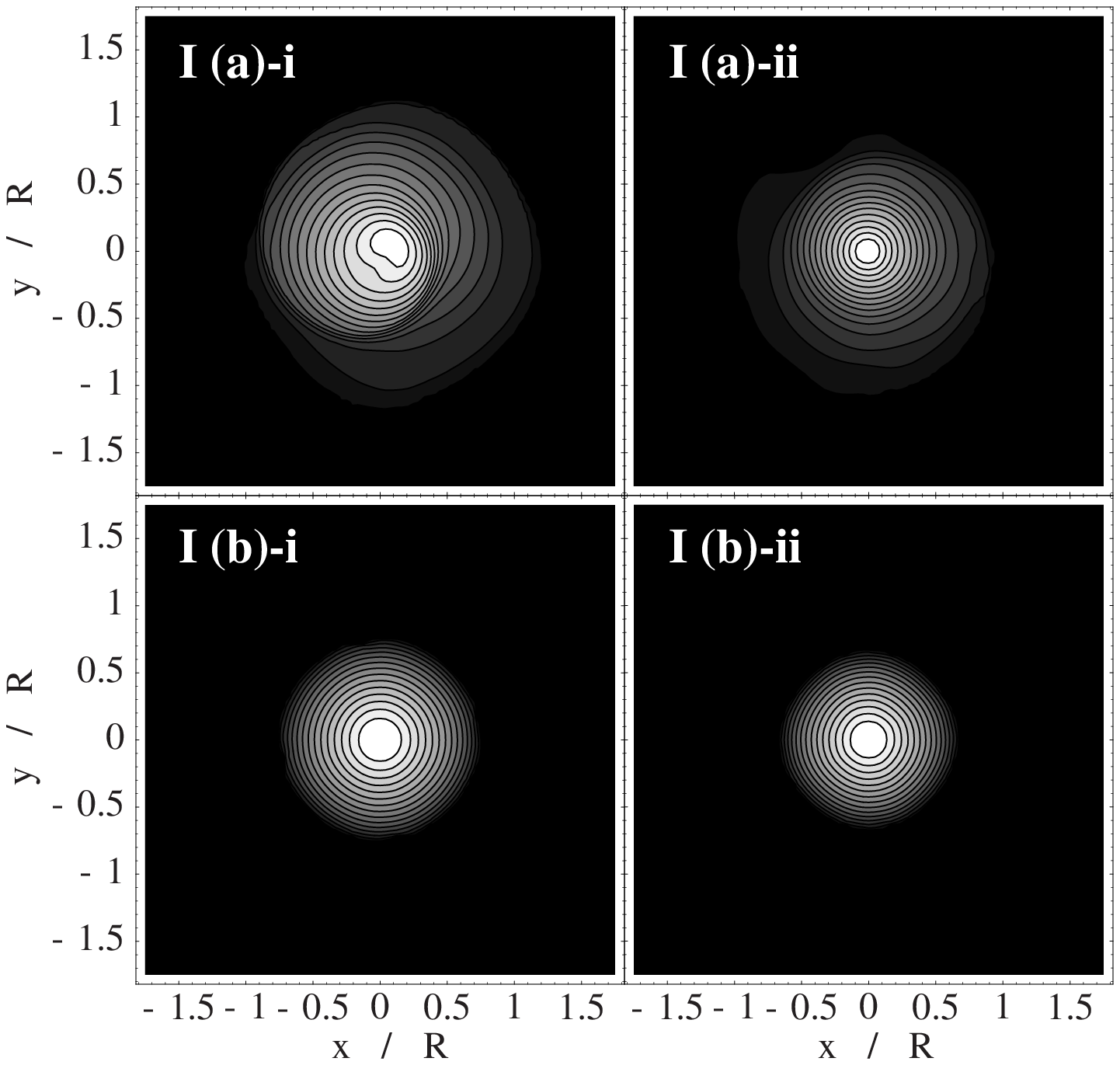}
\figcaption[f6.eps]{
Intermediate and final density contours in the equatorial plane for 
Model I (a) and Model I (b).  Snapshots are plotted at
($t/P_{\rm c}$, $\rho_{\rm max}/\rho_{\rm max}^{(0)}$, $d$) = 
(a)-i (16.3, 3.63, 0.287), and (b)-i (14.7, 2.08, 0.333), 
(a)-ii (23.3, 11.5, 0.287), and (b)-ii (20.6, 3.66, 0.333).
The contour lines denote densities 
$\rho/\rho_{\rm max} = 10^{- (16-i) d}  (i=1, \cdots, 15)$.
\label{fig:cttcon}}
\end{figure*}

We now focus on $m=1$ one-armed spiral instabilities.  Before we
analyze their dependence on the stiffness of the equation of state and
the degree of differential rotation in the following subsections, we
first want to reconfirm the findings of \citet{CNLB}.  To reconstruct
their initial data, we adopt a polytropic index of $n=3.33$ and a high
degree of differential rotation ($d/R_{e} = 0.2$).  We study two
different models, which are detailed in Table \ref{tab:m1test}.  The
more rapidly rotating Model I (a) \citep[the case $T/|W|=0.14$ of][]
{CNLB} has a toroidal structure, while Model I (b) \citep[the case
$T/|W|=0.09$ of][]{CNLB} does not. Confirming the results of
\citet{CNLB}, we find that Model I (a) develops an $m=1$ instability,
while Model I (b) remains stable.

The different stability properties of the two models can be seen in
Fig.~\ref{fig:cttdip}, where we show both diagnostics $D$ and $Q$.
For Model I (b), both diagnostics remain very small, indicating
stability\footnote{The small growth of the $m=1$ mode in Model I (b)
is a numerical artifact triggered by the initial perturbation; the
absence of an exponential growth indicates that this is not an
instability.}, while for Model I (a) both diagnostics grow.  The
dipole diagnostic $D$, however, grows more strongly than the
quadrupole diagnostic $Q$, indicating that the $m=1$ mode is the
dominant unstable mode.  This is also evident in the density contours
in Fig.~\ref{fig:cttcon}, which clearly exhibit the one-armed spiral
in Model I (a) at intermediate times.  In all cases that we found
to be unstable to an $m=1$ mode, we simultaneously found a growing
$m=2$ mode.

In Fig.~\ref{fig:cttrho} we show the maximum density $\rho_{\rm max}$
as a function of time for both models.  Even for the stable Model I
(b) the central density slowly increases over the course of several
central rotation periods.  This slow growth is due to numerical and
artificial viscosity, which tends to decrease the degree of
differential rotation.  As a consequence, the angular velocity at the
center decreases, which also decreases the rotational support of the
matter at the center, and hence leads to a slow increase of the
central density, even for supposedly stable stars (see also
Fig.~\ref{fig:ctttp}).  This effect is a numerical artifact, although
viscosities in stars in nature would have a very similar effect.  For
the unstable Model I (b), however, we find a much more rapid increase
in the central density.  This enhanced increase is caused by the
growing spiral instability as it redistributes the matter in the star
and destroys the toroidal structure (compare Fig.~\ref{fig:ctttp}).

Unlike for bar formation, where the bar persists for many rotational
periods \citep[compare \S~\ref{subsec:m2test},][]{Brown,SSBS}, we find
that $D$ and $Q$ start decreasing immediately after reaching a maximum
(see Fig.~\ref{fig:cttdip}, note that the decrease in $D$ is not as
dramatic as the decrease in $Q$).  This is also evident in
Fig.~\ref{fig:cttcon}, where the density contours approach axisymmetry
at late times.  As the spiral arm propagates through the star, it
rearranges the density profile, eliminates the toroidal structure, and
ultimately leads to a new axisymmetric equilibrium configuration.

\begin{figure*}[b]
\figurenum{10}
\epsscale{1.40}
\plotone{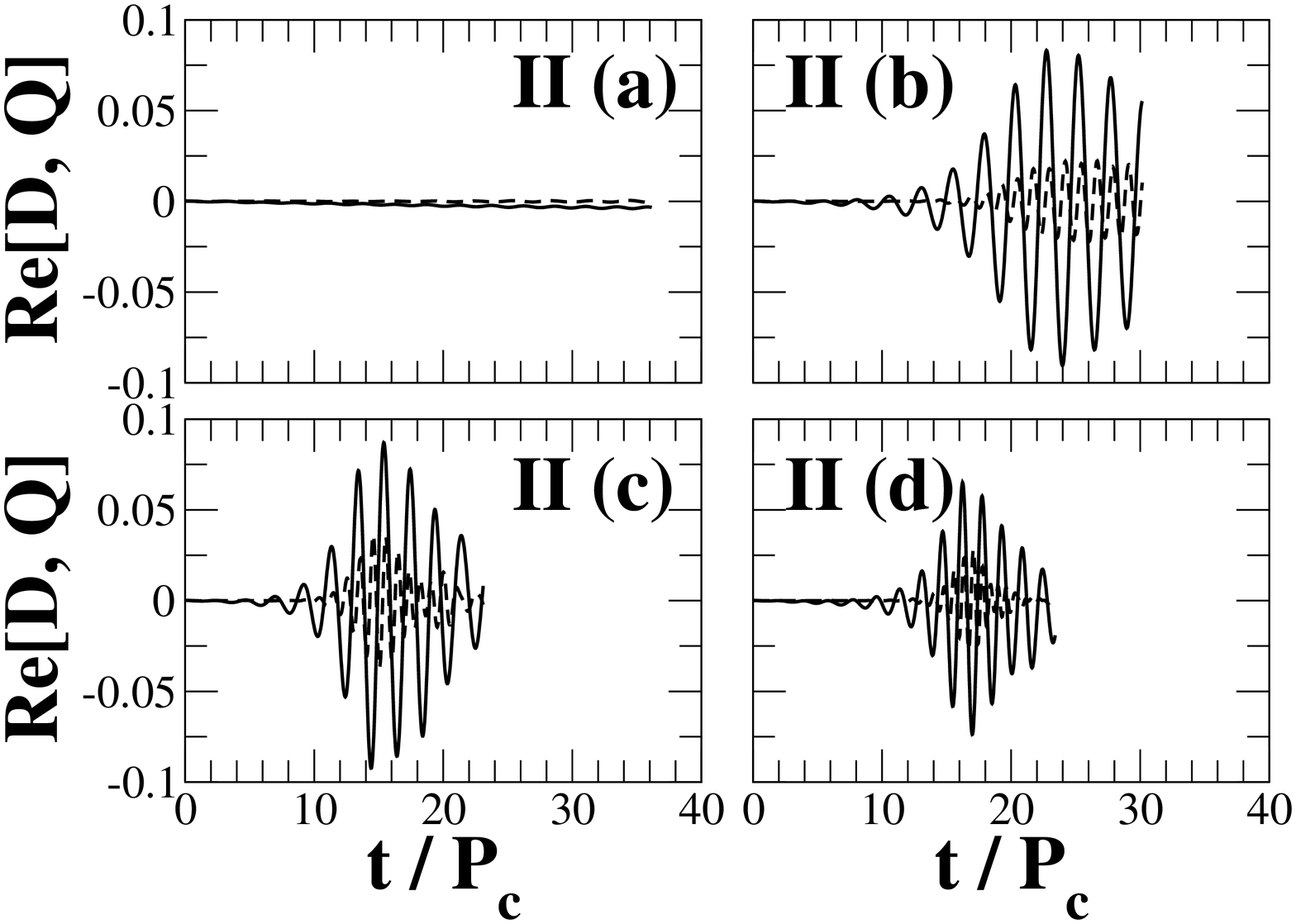}
\figcaption[f10.eps]{
Diagnostics $D$ and $Q$ as a function of $t/P_{\rm c}$ for Models II 
(see Table~\ref{tab:pindex}).  Solid and dotted lines denote $D$ 
and $Q$.  We terminate our simulation at 
$t \sim 25 P_{\rm c}$ or when the maximum density of the star exceeds 
about 10 times its initial value.
\label{fig:eosdip}}
\end{figure*}

\begin{center}
\begin{minipage}{6.5cm}
\epsfxsize 6.5cm
\epsffile{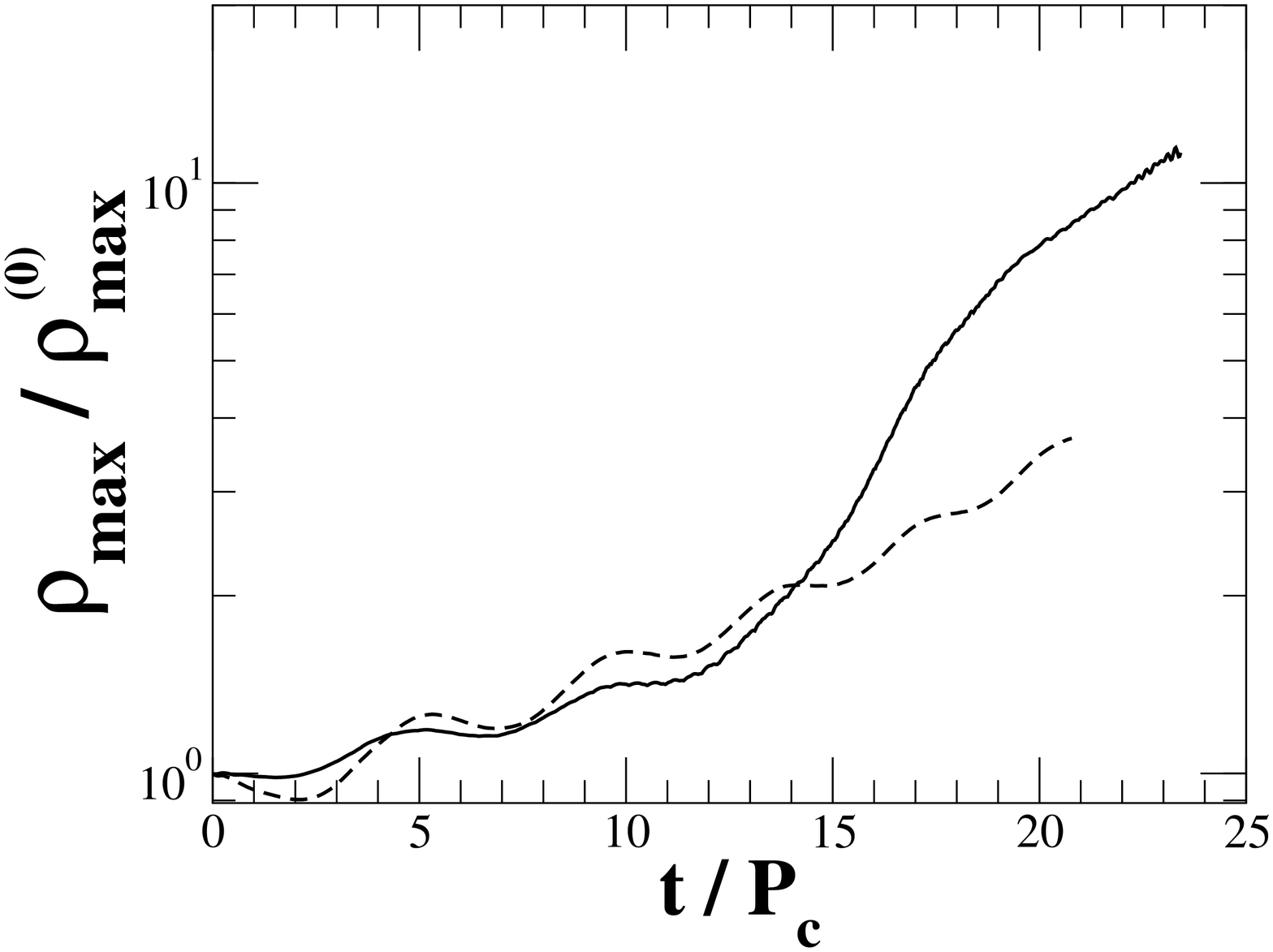}
\end{minipage}
\end{center}
\figurenum{7}
\figcaption[f7.eps]{
Maximum density $\rho_{\rm max}$ as a function of $t/P_{\rm c}$ for 
Model I (a) (solid) and Model I (b) (dotted).  We terminate our
simulation at $t \sim 20 P_{\rm c}$ or 
when the maximum density of the star exceeds about 10 times its
initial value $\rho^{(0)}_{\rm max}$.
\label{fig:cttrho}}
\vskip 12pt

\begin{center}
\begin{minipage}{6.5cm}
\epsfxsize 6.5cm
\epsffile{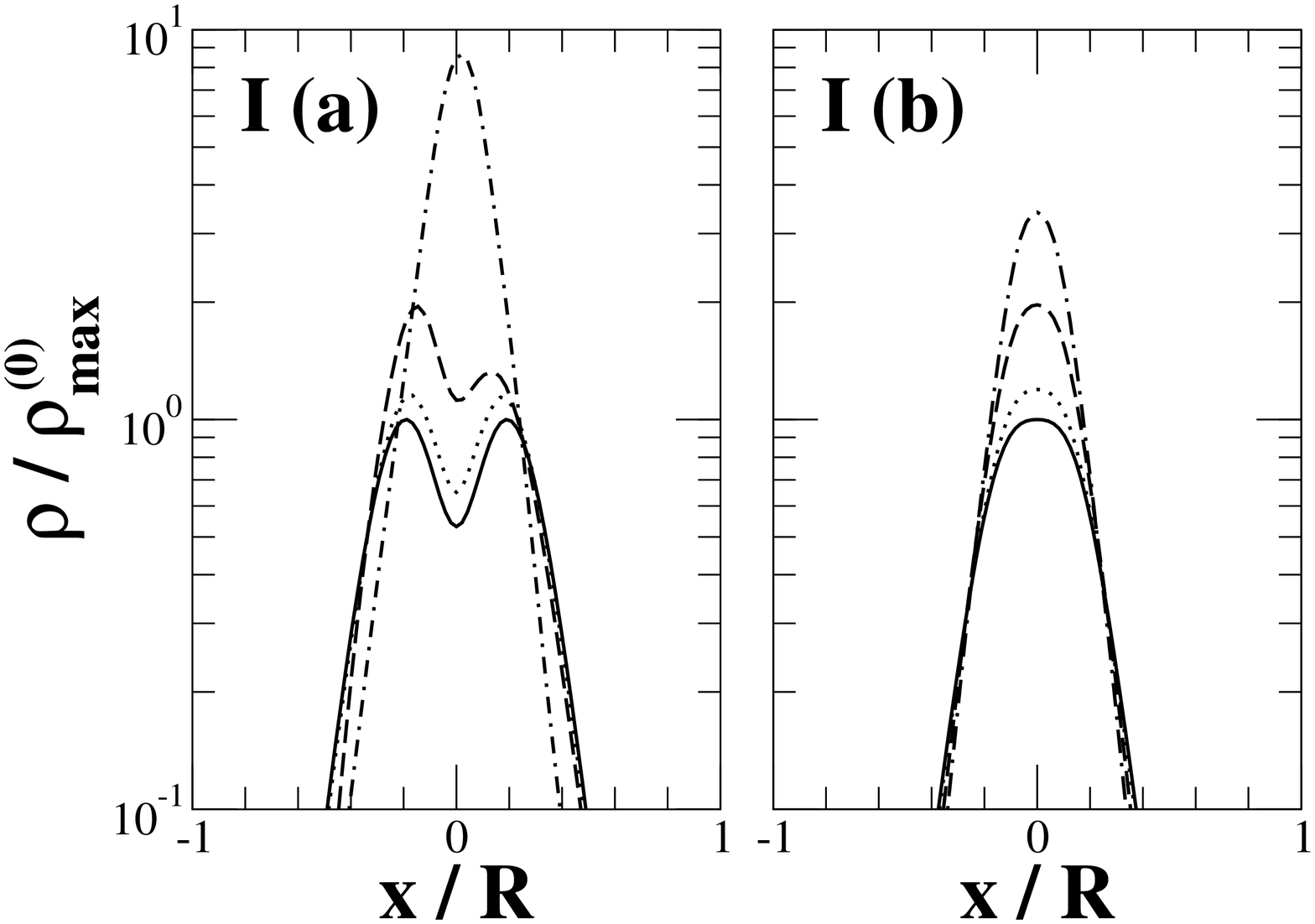}
\end{minipage}
\end{center}
\figurenum{8}
\figcaption[f8.eps]{
Density profiles along the $x$-axis during the
evolution for Models I (a) and I (b).  Solid, dotted, dashed, dash-dotted 
lines denote times
$t/P_{\rm c} =$ 
(a) ($1.16 \times 10^{-3}$, 6.99, 14.0, 21.0), 
(b) ($7.36 \times 10^{-4}$, 6.63, 13.3, 19.9), 
respectively.  Note that
the density distribution develops asymmetrically in the presence of
the $m=1$ mode instability, and that this instability destroys the
toroidal structure.
\label{fig:ctttp}}
\vskip 12pt

\begin{center}
\begin{minipage}{7.0cm}
\epsfxsize 7.0cm
\epsffile{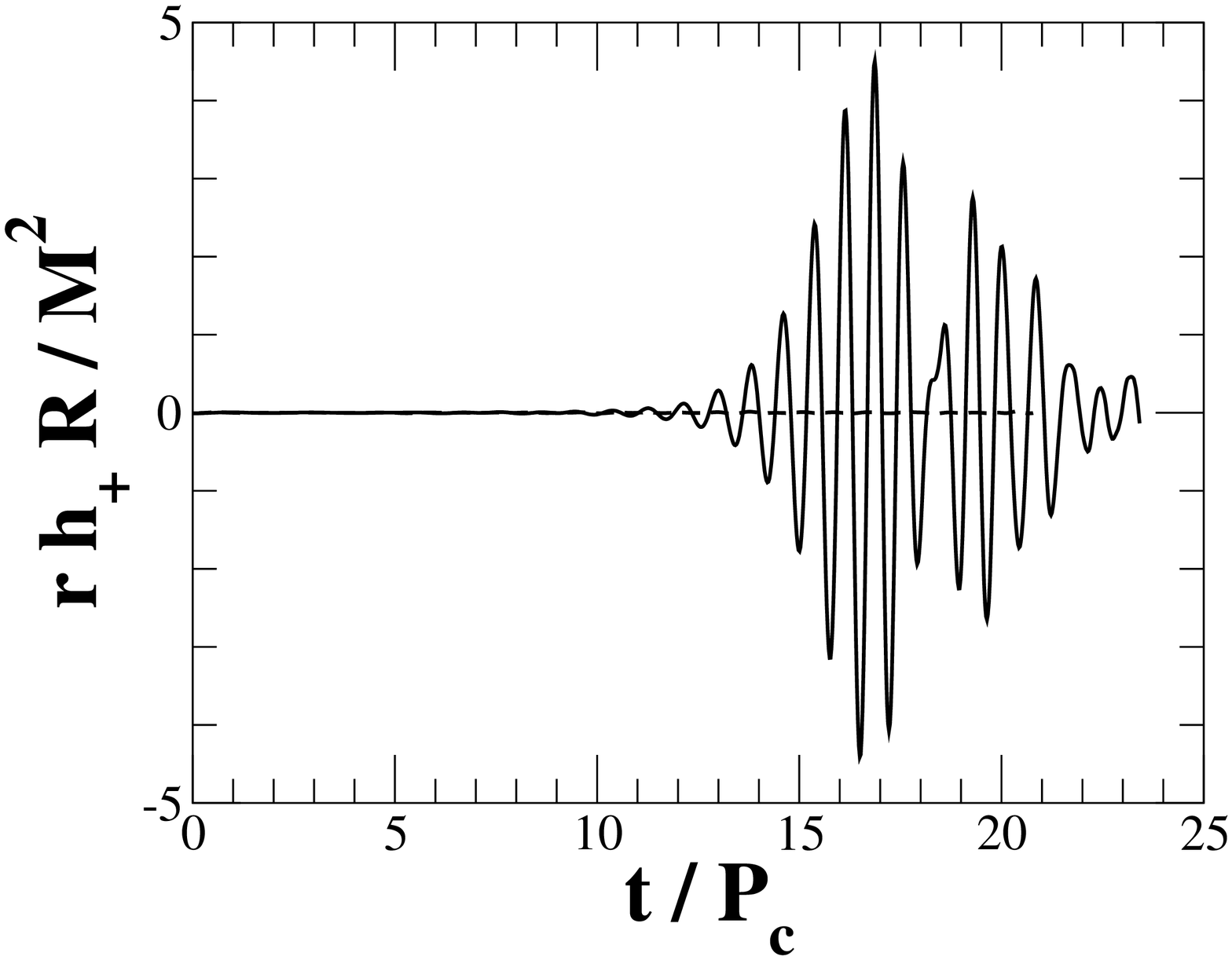}
\end{minipage}
\end{center}
\figurenum{9}
\figcaption[f9.eps]{
Gravitational waveforms as seen by a distant observer located on the 
$z$-axis for Model I (a) (solid line) and Model I (b) (dashed line).
\label{fig:cttgw}}
\vskip 12pt

In Fig.~\ref{fig:cttgw} we show the gravitational wave signal emitted
from this instability.  Gravitational radiation couples to quadrupole
moments, and the emitted radiation therefore scales with the
quadrupole diagnostic $Q$, which we always find excited along with the
$m=1$ instability.  We consistently find that the pattern period of
the the $m=2$ modes is very similar to that of the $m=1$ mode,
suggesting that the former is a harmonic of the latter (see Table
\ref{tab:Pperiod1}).  Since the diagnostic $Q$ does not remain at its
maximum amplitude after saturating, we find that the gravitational
wave amplitude is not nearly as persistent as for the bar mode
instability.  We also find that the gravitational wave period, here
$P_{\rm GW} \sim 0.7 P_{\rm c} \sim \Omega_{\rm c}^{-1}$, is different
from the value $P_{\rm GW} \sim 3.3 P_{\rm c} \sim \Omega_{\rm
eq}^{-1}$ we found for the bar mode in \S~\ref{subsec:m2test}, which
points to a difference in the generation mechanism.  Characteristic
wave frequencies $f_{\rm GW}$ correspond to the central rotation
period of the star.

\begin{figure*}
\figurenum{11}
\epsscale{1.40}
\plotone{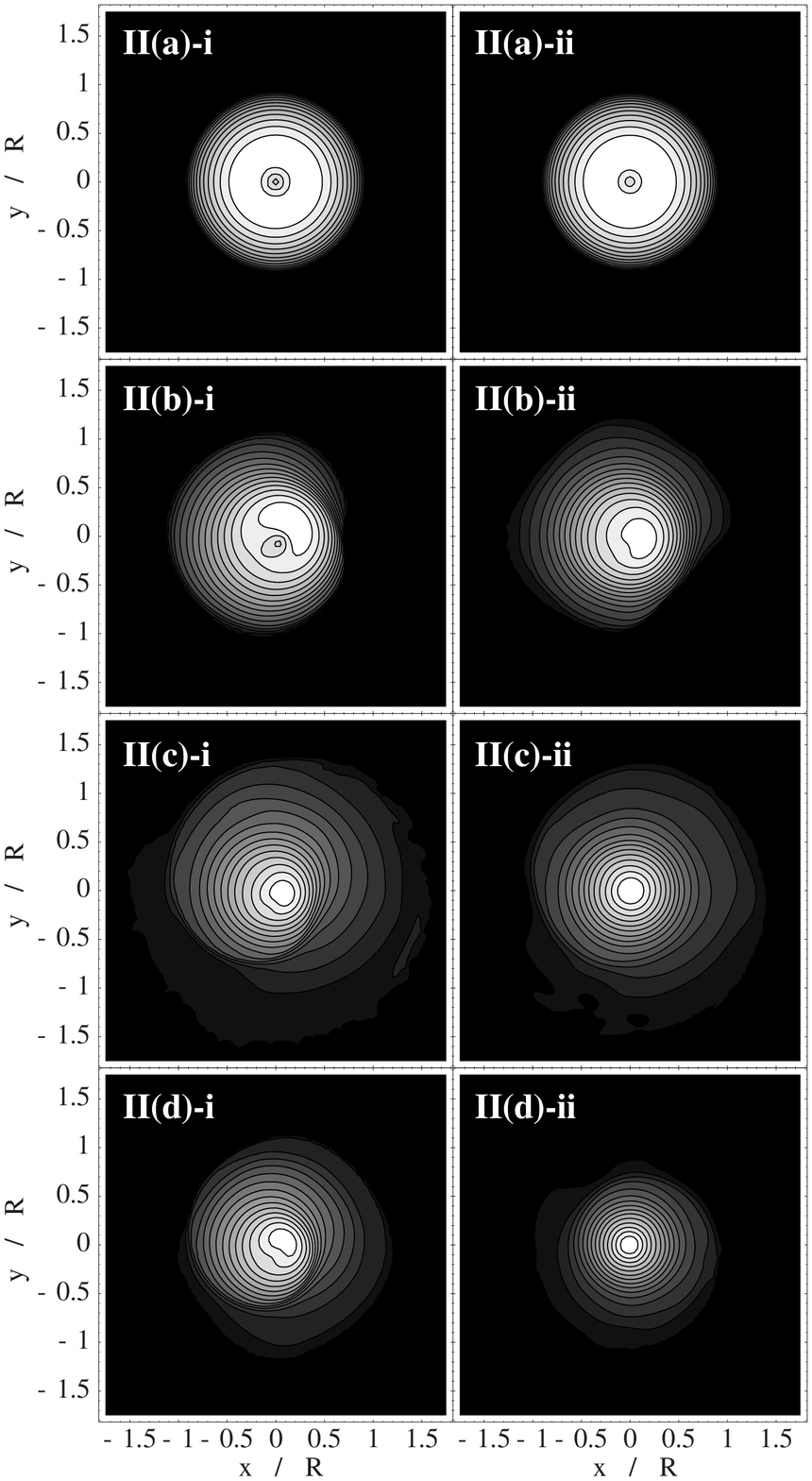}
\figcaption[f11.eps]{
Intermediate and final density contours in the equatorial plane 
for Models II.  Snapshots are plotted at
($t/P_{\rm c}$, $\rho_{\rm max}/\rho_{\rm max}^{(0)}$, $d$) = 
(a)-i (23.8, 1.14, 0.220), 
(b)-i (20.6, 1.90, 0.220), 
(c)-i (17.3, 4.98, 0.287), 
(d)-i (16.3, 3.63, 0.287), 
(a)-ii (34.7, 1.24, 0.220), 
(b)-ii (30.1, 2.92, 0.220), 
(c)-ii (25.2, 7.41, 0.287), and
(d)-ii (23.3, 11.5, 0.287). 
The contour lines denote densities 
$\rho/\rho_{\rm max} = 10^{- (16-i) d}  (i=1, \cdots, 15)$.
\label{fig:eoscon}}
\end{figure*}

\begin{table*}[t]
\begin{center}
\tablenum{3}
\label{tab:Pperiod1}
\centerline{\sc Table 3}
\centerline{\sc Comparison of Pattern Periods}
\vskip 6pt
\begin{tabular}{c c c c}
\hline
\hline
Model &
Period of $m=1$ mode [$P_{\rm c}$] &
Period of $m=2$ mode [$P_{\rm c}$] &
Pattern Period [$P_{\rm c}$]
\\
\hline
I (a) \tablenotemark{a} & $0.7$ & $1.6$ & $0.8$ \\
II (b) \tablenotemark{b} & $1.2$ & $2.5$ & $1.3$ \\
II (c) \tablenotemark{b} & $1.0$ & $2.0$ & $1.0$ \\
II (d) \tablenotemark{b} & $0.7$ & $1.6$ & $0.8$ \\
III (c) \tablenotemark{c} & $0.7$ & $1.6$ & $0.8$ \\
III (d) \tablenotemark{c} & $0.7$ & $1.6$ & $0.8$
\\
\hline
\hline
\end{tabular}
\vskip 12pt
\begin{minipage}{13cm}
{${}^{a}$ {See Fig. \ref{fig:cttdip}}}\\
{${}^{b}$ {See Fig. \ref{fig:eosdip}}}\\
{${}^{c}$ {See Fig. \ref{fig:ddrdip}}}
\end{minipage}
\end{center}
\end{table*}

\begin{figure*}[b]
\figurenum{13}
\epsscale{1.40}
\plotone{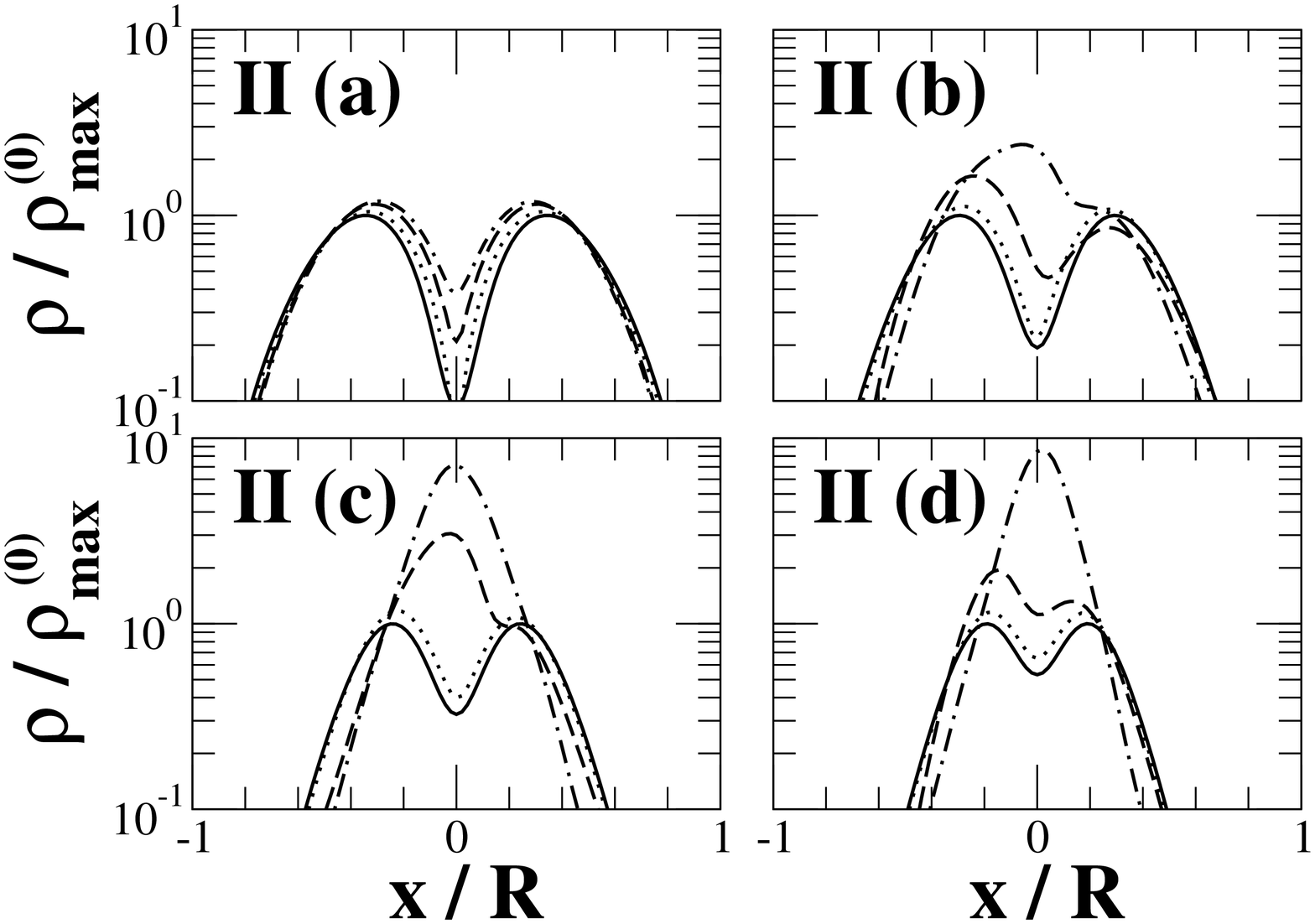}
\figcaption[f13.eps]{
Density profiles along the $x$-axis during the evolution of Models II.  
Solid, 
dotted, dashed, dash-dotted line denote at time $t / P_{\rm c} =$ 
(a) ($2.17 \times 10^{-3}$, 10.8, 21.7, 32.5),
(b) ($1.56 \times 10^{-3}$, 9.56, 19.0, 28.5),
(c) ($1.33 \times 10^{-3}$, 7.97, 15.9, 23.9),
(d) ($1.16 \times 10^{-3}$, 6.98, 14.0, 21.0),
respectively. Note that the toroidal structure vanishes at late
times for Models II (b), II (c), and II (d). 
\label{fig:eostp}}
\end{figure*}

\begin{center}
\begin{minipage}{7.0cm}
\epsfxsize 7.0cm
\epsffile{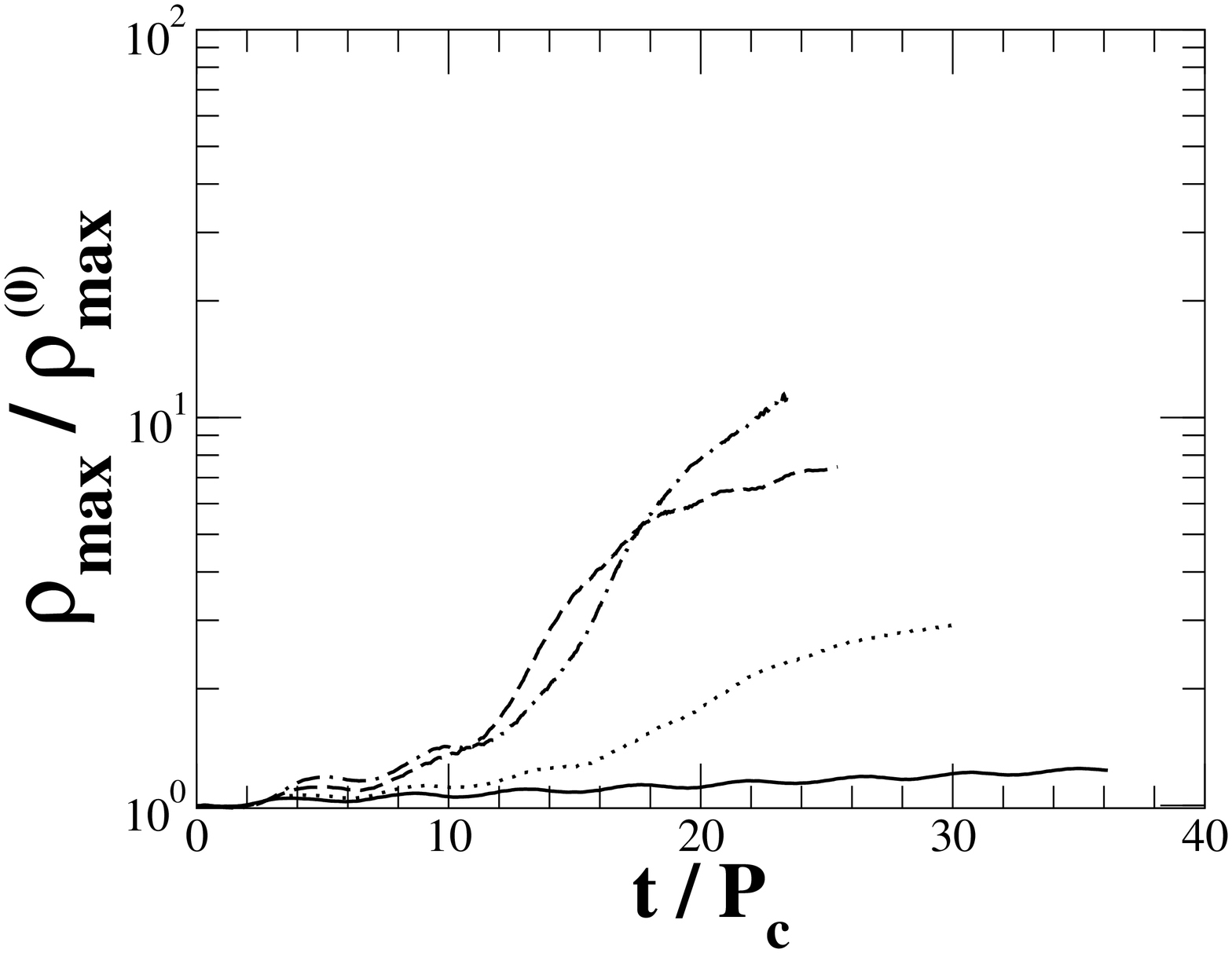}
\end{minipage}
\end{center}
\figurenum{12}
\figcaption[f12.eps]{
Maximum density $\rho_{\rm max}$ as a function of $t/P_{\rm c}$ for 
Model II (a) (solid line), Model II (b) (dotted line), Model II (c)
(dashed line), and Model II (d) (dash-dotted line).
\label{fig:eosrho}}
\vskip 12pt

\begin{figure*}
\figurenum{14}
\epsscale{1.40}
\plotone{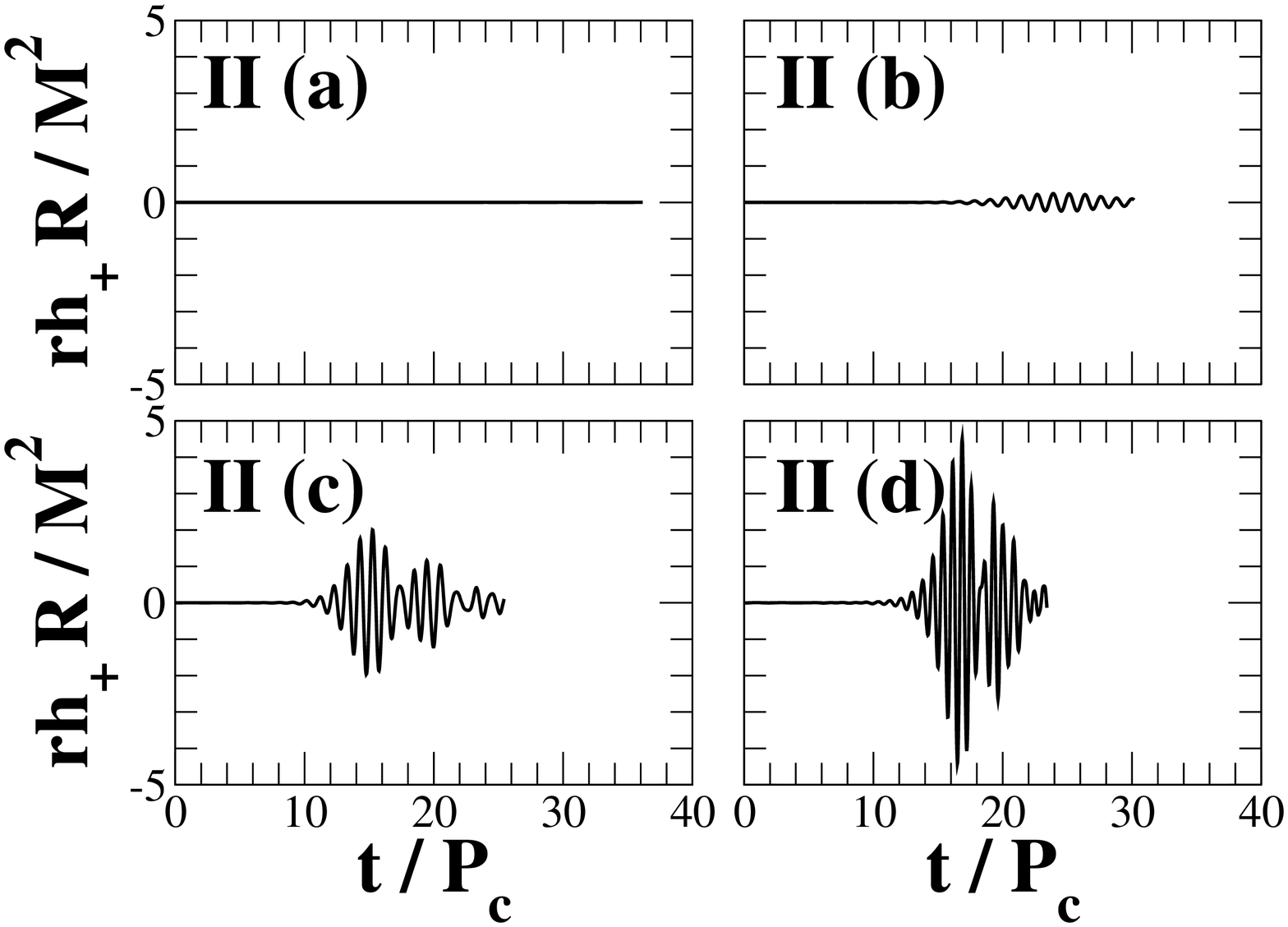}
\figcaption[f13.eps]{
Gravitational waveforms as seen by a distant observer located on the 
$z$-axis for Models II.
\label{fig:eosgw}}
\end{figure*}

\begin{figure*}
\figurenum{15}
\epsscale{1.40}
\plotone{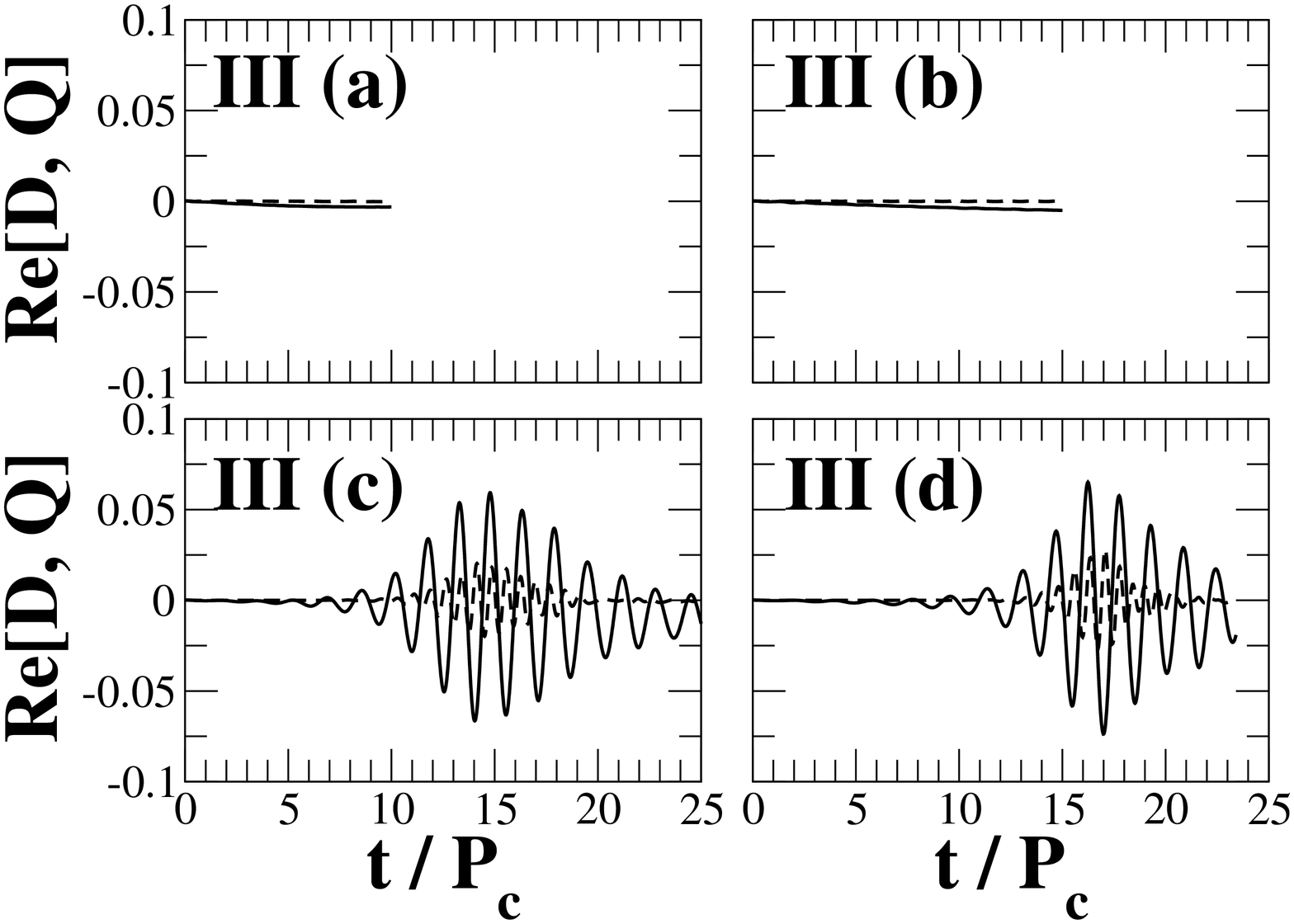}
\figcaption[f15.eps]{
Diagnostics $D$ and $Q$ as a function of $t/P_{\rm c}$ for Models III 
(see Table \ref{tab:drot}). Solid and dotted lines denote $D$ 
and $Q$.  
\label{fig:ddrdip}}
\end{figure*}

\begin{figure*}
\figurenum{16}
\epsscale{1.40}
\plotone{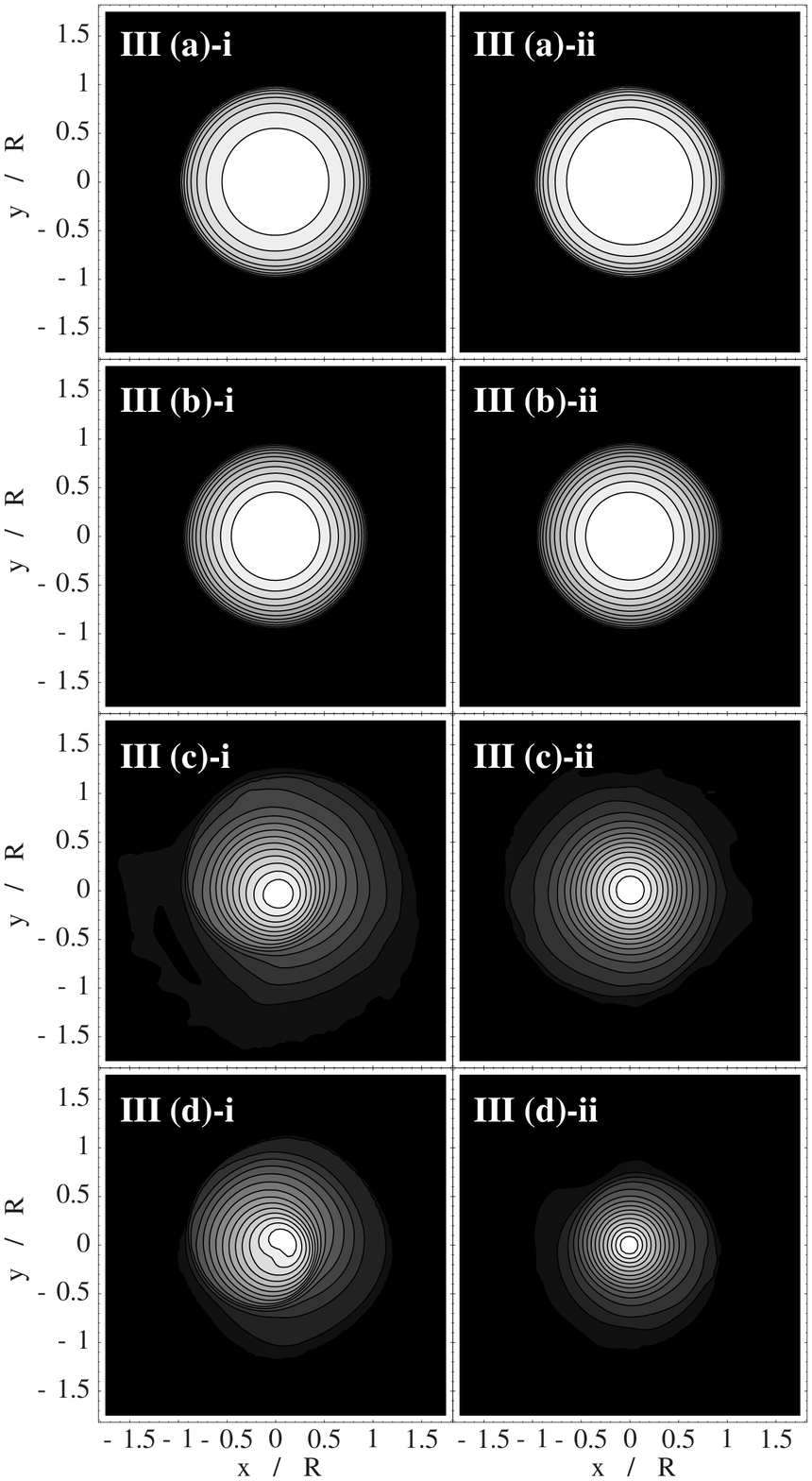}
\figcaption[f16.eps]{
Intermediate and final density contours in the equatorial plane 
for Models III.  Snapshots are plotted at
($t/P_{\rm c}$, $\rho_{\rm max}/\rho_{\rm max}^{(0)}$, $d$) = 
(a)-i (6.81, 1.01, 0.200), 
(b)-i (10.4, 1.07, 0.267), 
(c)-i (17.6, 3.72, 0.287), 
(d)-i (16.3, 3.63, 0.287), 
(a)-ii (9.90, 1.02, 0.200), 
(b)-ii (14.8, 1.09, 0.267), 
(c)-ii (25.6, 5.05, 0.287), and 
(d)-ii (23.3, 11.5, 0.287). 
The contour lines denote densities 
$\rho = \rho_{\rm max} \times 10^{- (16-i) d}  (i=1, \cdots, 15)$.
\label{fig:ddrcon}}
\end{figure*}

\begin{table*}[t]
\begin{center}
\tablenum{4}
\label{tab:pindex}
\centerline{\sc Table 4}
\centerline{\sc Initial data sequence varying the polytropic index.}
\vskip 6pt
\begin{tabular}{c c c c c c c c c}
\hline
\hline
Model &
$n$\tablenotemark{a} &
$d / R_{\rm eq}$ &
$R_{\rm p} / R_{\rm eq}$ &
$\Omega_{\rm c} / \Omega_{\rm eq}$ &
$\rho_{\rm c} / \rho_{\rm max}$ &
$R_{\rm maxd}/R_{\rm eq}$ &
$T/|W|$ &
$m=1$ stability
\\
\hline
II (a) & $2.00$ & $0.20$ & $0.271$ & $26.0$ & $0.091$  & $0.349$ & 
$0.145$ & Stable
\\
II (b) & $2.50$ & $0.20$ & $0.354$ & $26.0$ & $0.193$  & $0.295$ & 
$0.145$ & Unstable
\\
II (c) & $3.00$ & $0.20$ & $0.396$ & $26.0$ & $0.325$  & $0.243$& 
$0.147$ & Unstable
\\
II (d)\tablenotemark{b} & $3.33$ & 
$0.20$ & $0.417$ & $26.0$ & $0.531$ & $0.192$ &
$0.144$ & Unstable
\\
\hline
\hline
\end{tabular}
\vskip 12pt
\begin{minipage}{13cm}
{${}^{a}$ {polytropic index}}\\
{${}^{b}$ {Same as Model I (a).}}
\end{minipage}
\end{center}
\end{table*}

\begin{table*}[b]
\begin{center}
\tablenum{5}
\label{tab:drot}
\centerline{\sc Table 5}
\centerline{\sc Initial data sequence varying the degree of differential 
rotation.}
\vskip 6pt
\begin{tabular}{c c c c c c c c c}
\hline
\hline
Model &
$n$ &
$d / R_{\rm eq}$ &
$R_{\rm p} / R_{\rm eq}$ &
$\Omega_{\rm c} / \Omega_{\rm eq}$ &
$\rho_{\rm c} / \rho_{\rm max}$ &
$R_{\rm maxd}/R_{\rm eq}$ &
$T/|W|$ &
$m=1$ stability
\\
\hline
III (a) & $1.00$ & $0.62$ & $0.500$ & $3.60$ & $0.992$ & $0.189$ & 
$0.150$ & Stable
\\
III (b) & $2.00$ & $0.41$ & $0.479$ & $6.95$ & $0.935$  & $0.198$ & 
$0.150$ & Stable
\\
III (c) & $3.00$ & $0.25$ & $0.438$ & $17.0$ & $0.695$  & $0.197$& 
$0.147$ & Unstable
\\
III (d) \tablenotemark{a} & $3.33$ & $0.20$ & $0.417$ & $26.0$ & $0.531$ &
$0.192$ &
$0.144$ & Unstable
\\
\hline
\hline
\end{tabular}
\vskip 12pt
\begin{minipage}{13cm}
{${}^{a}$ {Model I (a) in Table. \ref{tab:m1test}.}}
\end{minipage}
\end{center}
\end{table*}

The results of this Subsection confirm the findings of \citet{CNLB},
and establish that stars with soft equations of state and large degrees
of differential rotation are unstable to one-armed spiral arm
formation.  Such stars have a toroidal structure which is erased by
the growing $m=1$ mode.  One might be lead to believe that this
toroidal structure is a necessary and perhaps even a sufficient
condition for the growth of the $m=1$ instability.  In the following
two Subsections we analyze the dependence of the onset of instability
on both the stiffness of the equation of state and the degree of
differential rotation, and find that toroidal structure alone is
not sufficient for a one-armed spiral instability.

\subsection{Stiffness of the equation of state}
\label{subsec:eosdip}
We parameterize the stiffness of the equation of state by varying the
polytropic index $n$ between $n = 3.33$ and $n = 2$.  In this sequence
we keep the degree of differential rotation (i.e.~$d$ and hence
$\Omega_{\rm c}/\Omega_{\rm eq}$) fixed, and adjust the overall
rotation rate (parameterized by $R_{\rm p}/R_{\rm eq}$) so that the
value of $T/|W|$ remains very close to 0.144 (as for Model I [a]).  We list
our four different Models II in Table \ref{tab:pindex}, and note that
Model II (d) is identical to Model I (a).  

Figure~\ref{fig:eosdip}, where we plot the dipole diagnostic $D$ as a
function of time, clearly shows that an $m=1$ instability is excited
in Models II (b) and II (c) in addition to Model II (d).  After
reaching saturation, $D$ decreases again, similar to Model I (a) which
we described in detail in \S~\ref{subsec:m1test}.  Model II (a),
however, which has the most pronounced toroidal structure, remains
stable.  These findings are also evident in Fig.~\ref{fig:eoscon},
where we show density contours of intermediate and final
configurations.

In Fig.~\ref{fig:eosrho}, we show the maximum density as a function of
time.  As we have seen in \S~\ref{subsec:m1test}, the maximum density
slowly increases in all cases due to dissipation of differential
rotation.  Once the one-armed spiral forms in Models II (b) and II
(c), however, this increase is much more rapid, which indicates again
that the unstable mode rearranges the matter in the star and destroys
the toroidal structure.  This effect can also be seen in the density
profiles in Fig.~\ref{fig:eostp}.

We show the gravitational wave signal emitted from Models II in
Fig.~\ref{fig:eosgw}.  As we found in \S~\ref{subsec:m1test}, and
consistent with the diagnostics $D$ and $Q$, the gravitational wave
signal emitted by the one-armed spiral mode does not persist over many
rotational periods, and instead decays fairly rapidly after it has
been excited.  This characteristic is very different from what has
been found for $m=2$ bar mode instabilities \citep[compare
\S~\ref{subsec:m2test} and][]{Brown,SSBS}.  We also find that the
maximum wave amplitude is much smaller than can be found for
configurations unstable to a pure bar mode (compare
Fig.~\ref{fig:bargw}) as gravitational radiation requires a quadrupole
distortion and the $m=2$ perturbation in Models II is only being
excited as a lower-amplitude harmonic of the $m=1$ mode (as 
suggested by a comparison of the pattern periods; see 
Table \ref{tab:Pperiod1}).

\begin{center}
\begin{minipage}{7.0cm}
\epsfxsize 7.0cm
\epsffile{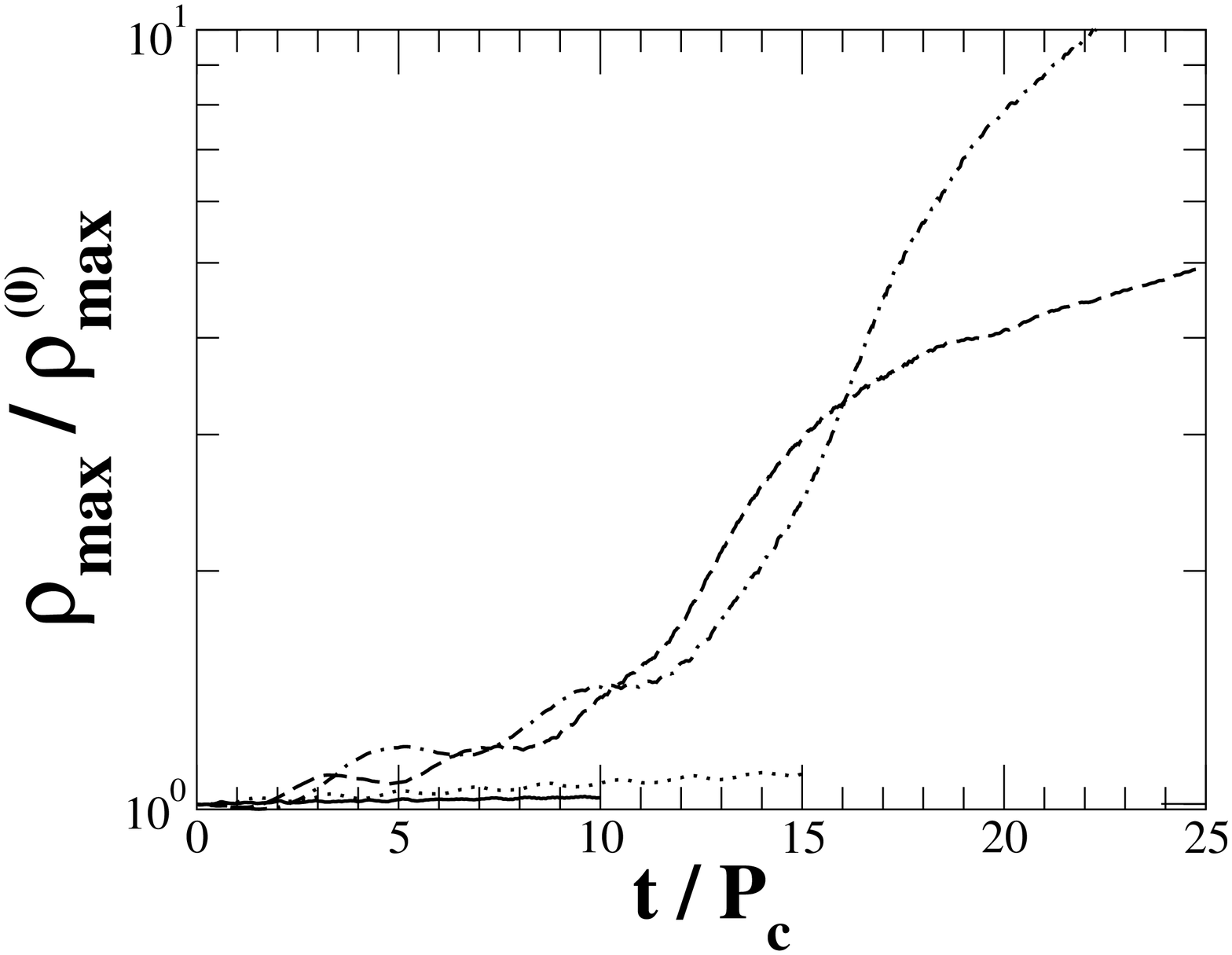}
\end{minipage}
\end{center}
\figurenum{17}
\figcaption[f17.eps]{
Maximum density $\rho_{\rm max}$ as a function of $t/P_{\rm c}$ for 
Model III (a) (solid line), Model III (b) (dotted line), Model III (c) 
(dashed line), and Model III (d) (dash-dotted line).
\label{fig:ddrrho}}
\vskip 12pt

\begin{figure*}
\figurenum{18}
\epsscale{1.40}
\plotone{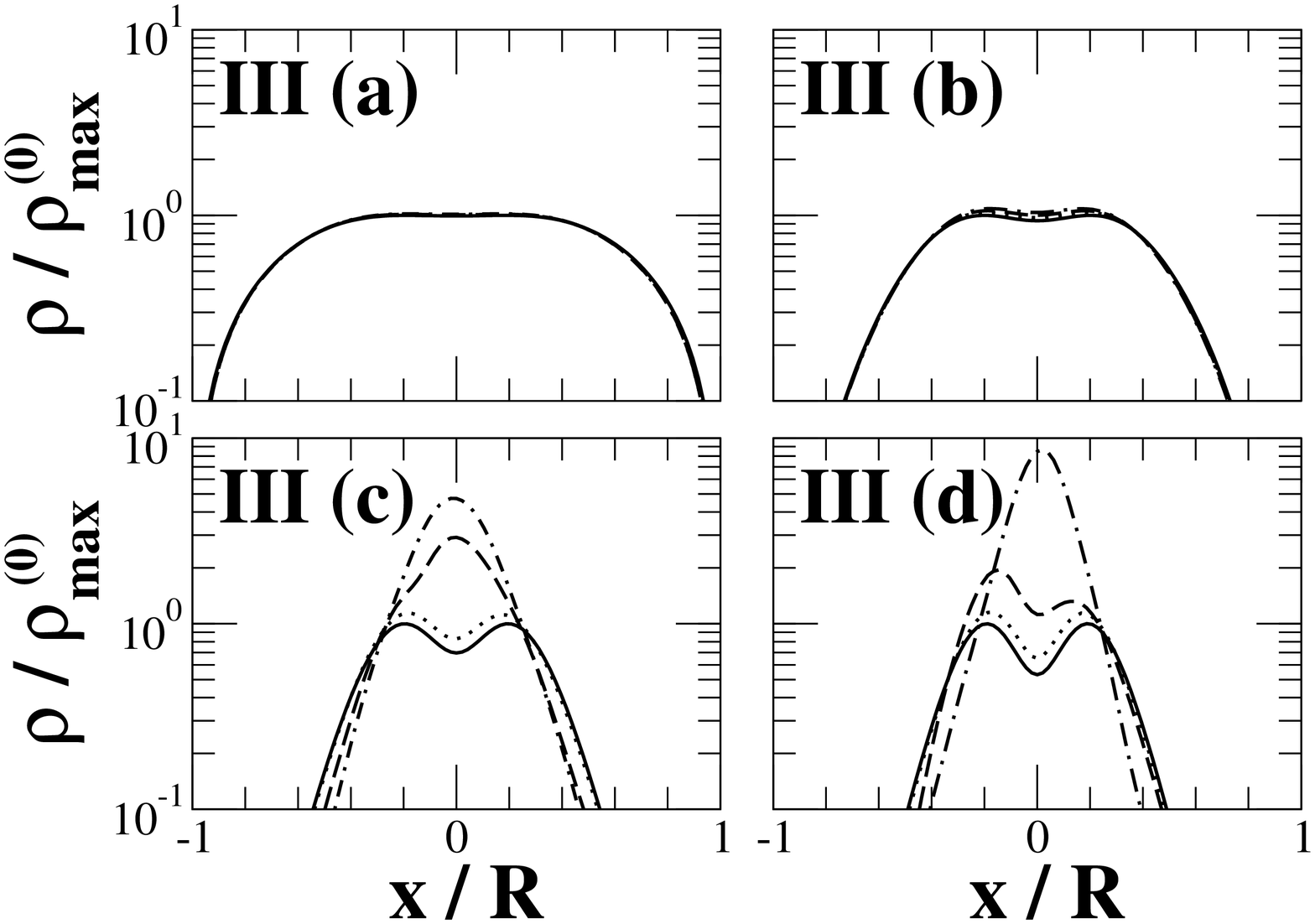}
\figcaption[f18.eps]{
Density profiles along the $x$-axis during the evolution for Models III.  
Solid, dotted, dashed, dash-dotted line denote at time $t/P_{\rm c} =$ 
(a) ($3.09 \times 10^{-4}$, 3.09, 6.19, 9.28),
(b) ($4.95 \times 10^{-4}$, 4.95, 9.89, 14.8),
(c) ($8.82 \times 10^{-4}$, 8.82, 17.6, 26.4),
(d) ($1.16 \times 10^{-3}$, 6.99, 14.0, 21.0),
respectively. Note that the toroidal structure vanishes at the late 
time in Models III (c) and III (d). 
\label{fig:ddrtp}}
\end{figure*}

\begin{figure*}
\figurenum{19}
\epsscale{1.40}
\plotone{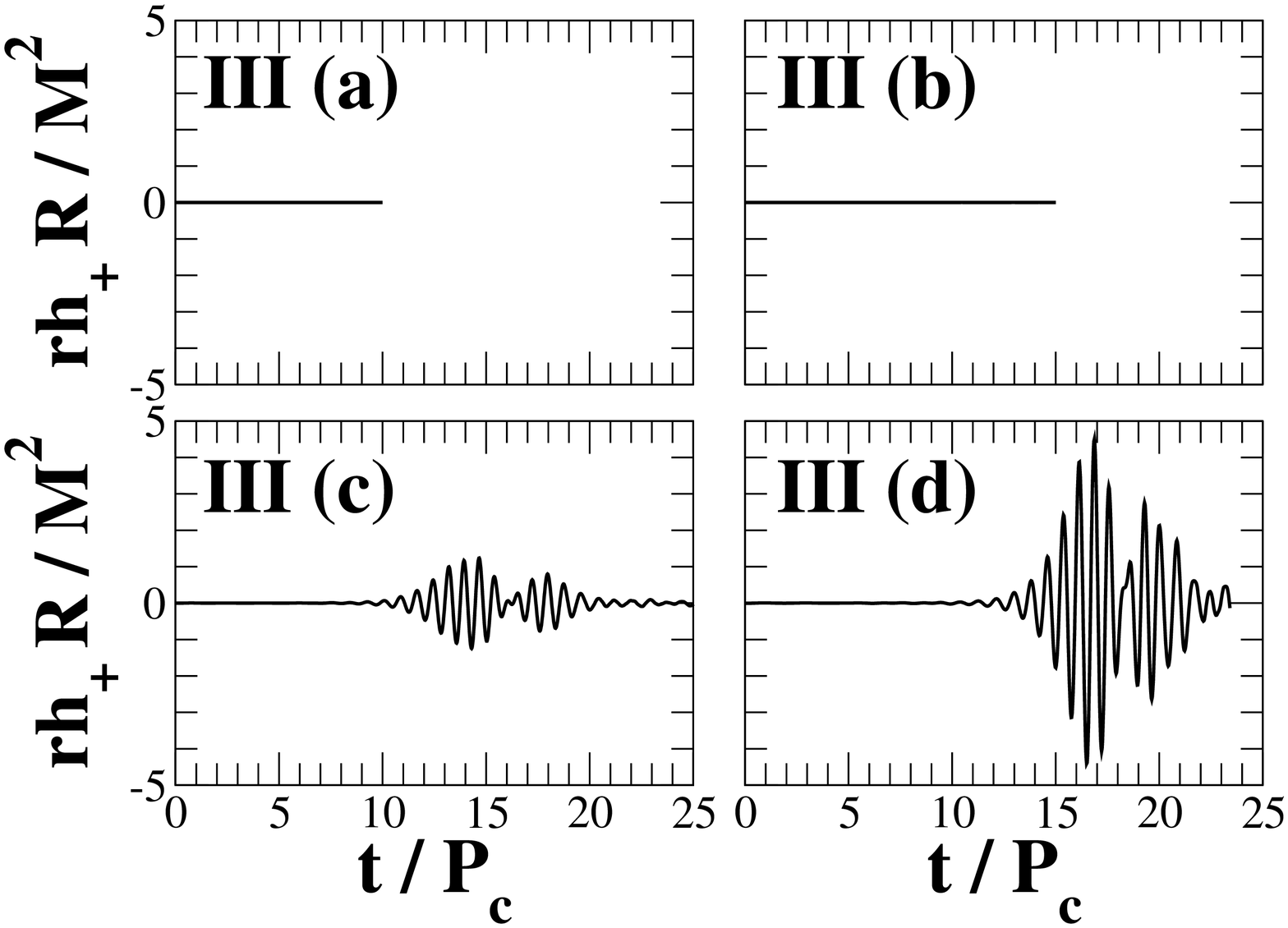}
\figcaption[f19.eps]{
Gravitational waveforms as seen by a distant observer located on the 
$z$-axis for Models III.
\label{fig:ddrgw}}
\end{figure*}

\subsection{Degree of differential rotation}

We now focus on the dependence of the one-armed spiral instability
on the degree of differential rotation.  Starting again with Model I (a),
we now increase the parameter $d$ to explore more modest degrees of
differential rotations.  As before, we would like to keep $\beta \sim
0.14$ in this sequence.  For very soft equations of state, this value
can only be achieved for very strong degrees of differential rotation.
Therefore, in order to keep $\beta$ approximately constant, we
simultaneously have to decrease $n$ as we decrease the degree of
differential rotation\footnote{This means that our results do not
separate the dependence on the degree of differential rotation from
the dependence on the stiffness of the equation of state as cleanly as
one might wish.}.  We list the details of our Models III in Table
\ref{tab:drot}.

We show the dipole diagnostic $D$ as a function of time in
Fig.~\ref{fig:ddrdip}, which shows that Models III (a) and III (b) are
stable against one-armed spiral formation while Models III (c) and III
(d) (which is the same as Model I [a]) are not.  The same conclusion
can be drawn from the density snapshots in Fig.~\ref{fig:ddrcon}.  As
in \S~\ref{subsec:eosdip}, we find that the one-armed spiral results
in a large increase in the central density (Fig.~\ref{fig:ddrrho}),
and an elimination of the toroidal structure (Fig.~\ref{fig:ddrtp}).
\citet{TH} similarly found that the elimination of the toroidal structure
is related to an outward transport of angular momentum.  To quantify
this effect, we monitored the angular momentum distribution in Model III(d)
by computing a mean radius of angular momentum
\begin{eqnarray}
<j> = \frac{M [\int dv \rho \sqrt{x^2 + y^2} (x v^y - y v^x)]}
{J [\int dv \rho \sqrt{x^2 + y^2}]}.
\end{eqnarray}
Initially this mean radius is 1.1, but increases to 1.5 at $t=24.5
P_{\rm c}$, indicating that in fact the $m=1$ mode transports angular
momentum outward.

We show gravitational waveforms from Models III in
Fig.~\ref{fig:ddrgw}.  We again find that the amplitude decreases
after reaching a maximum, which is a typical behavior of $m=1$ 
instability (see Table \ref{tab:Pperiod1} that the pattern period of 
diagnostics $D$ and $Q$ are the same).  In some cases, however, this 
decrease is not monotonic, and the amplitude may increase again to form 
several distinct wave packets.  Our numerical data are not sufficient to
determine the generic character of the gravitational waves emitted
from $m=1$ instabilities, and we expect that this will be subject of
future investigations.  The problem is that the growth of central
concentration during the evolution exceeds the ability of our code to
resolve the innermost regions for arbitrary long times in all cases.

\section{Discussion}
\label{sec:Discussion}

We have studied the conditions under which Newtonian, differentially
rotating stars are dynamically unstable to an $m=1$ one-armed spiral
instability, and found that both soft equations of state and a high
degree of differential rotation are necessary to trigger the
instability.  For sufficiently soft equations of state and
sufficiently high degrees of differential rotation we found that stars
are dynamically unstable even at the small values of $T/|W| \sim 0.14$
considered in this paper.

While we find that a toroidal structure alone is not sufficient for
the $m=1$ instability, all the models that are unstable do have a
toroidal structure, suggesting that this may be a necessary condition.
The growing $m=1$ mode redistributes both matter and angular momentum
in the unstable star and destroys the toroidal structure after a few
central rotation periods.

Quasi-periodic gravitational waves emitted by stars with $m=1$
instabilities have smaller amplitudes than those emitted by stars
unstable to the $m=2$ bar mode.  For $m=1$ modes, the gravitational
radiation is emitted not by the primary mode itself, but by the $m=2$
secondary harmonic which is simultaneously excited, albeit at a lower
amplitude (see Fig.~\ref{fig:cttdip}).  Unlike the case for
bar-unstable stars, the gravitational wave signal does not persist of
many periods, but instead is damped fairly rapidly in most of the
cases we have examined.

We have plotted typical wave forms for stars unstable to $m=2$ bar
modes in Fig.~\ref{fig:bargw} and for stars unstable to one-armed
spiral $m=1$ modes in Figs.~\ref{fig:cttgw}, \ref{fig:eosgw} and
\ref{fig:ddrgw}.  Characteristic wave frequencies $f_{\rm GW}$ are
seen to be $\sim P^{-1}_{\rm c} \sim \Omega_{\rm c}$, and are
considerably higher than $\Omega_{\rm eq} \lesssim (M/R^3)^{1/2}$ due
to appreciable differential rotation.  For supermassive stars ($M
\gtrsim 10^5 M_{\odot}$) the amplitudes and frequencies of these waves
fall well within the detectable range of LISA (see, e.g., \citet{NS}).

\acknowledgments
We thank an anonymous referee for the careful reading of this
manuscript and many constructive suggestions.  This work was supported
by JSPS Grant-in-Aid for young scientists (No. 1400927), NSF Grants
PHY-0090310 and PHY-0205155 and NASA Grant NAG 5-10781 at the
University of Illinois at Urbana-Champaign and NSF Grant PHY-0139907
at Bowdoin College.  Numerical computations were performed on the NEC
SX-5 machine in the Yukawa Institute for Theoretical Physics, Kyoto
University, on the VPP-800 machine in the Academic Center for
Computing and Media Studies, Kyoto University, and on the VPP-5000
machine in the Astronomical Data Analysis Center, National
Astronomical Observatory of Japan.  MS thanks Department of Physics,
University of Illinois at Urbana-Champaign for their hospitality.


\end{document}